\documentclass[Times,twocolumn,tighten,deluxetable]{aastex7}

\usepackage{graphicx}
\usepackage{rotating}
\usepackage{lineno}
\usepackage{multirow}
\usepackage{enumerate}
\submitjournal{ApJ}

\shorttitle{Accretion Flow Properties of Swift J1727.8-1613 and Estimation of the BH Mass}
\shortauthors{Debnath et al.}

\begin{document}

%\title{Evolution of QPO during Rising Phase of Discovery Outburst of Swift J1727.8-1613: Estimation of Mass from Spectro-Temporal Study}
\title{Accretion Flow Properties of Swift J1727.8-1613 During Its Rising Phase of the Discovery Outburst: Estimation of Mass from Spectro-Temporal Study}

\correspondingauthor{Dipak Debnath}
\email{dipakcsp@gmail.com}

\author[0000-0003-1856-5504]{Dipak Debnath}
\affiliation{Institute of Astronomy, National Tsing Hua University, Hsinchu 300044, Taiwan}
\affiliation{Institute of Astronomy Space and Earth Science, P 177, CIT Road, Scheme 7m, Kolkata 700054, India}
\email{dipakcsp@gmail.com}

\author[0000-0002-5617-3117]{Hsiang-Kuang Chang}
\affiliation{Institute of Astronomy, National Tsing Hua University, Hsinchu 300044, Taiwan}
\affiliation{Department of Physics, National Tsing Hua University, Hsinchu 300044, Taiwan}
\email{hkchang@mx.nthu.edu.tw}

%\author{Anuj Nandi}
%\affiliation{Space Astronomy Group, ISITE Campus, U R Rao Satellite Centre, Bengaluru, 560037, India}

\author[0000-0002-6640-0301]{Sujoy Kumar Nath}
\affiliation{Institute of Astronomy, National Tsing Hua University, Hsinchu 300044, Taiwan}
\affiliation{Indian Center for Space Physics,  466 Barakhola, Netai Nagar, Kolkata 700099, India}
\email{sujoynath0007@gmail.com}

\author[0000-0002-9998-7591]{Lev Titarchuk}
\affiliation{Dipartimento di Fisica, University di Ferrara, Via Saragat 1, I-44122 Ferrara, Italy}
\email{titarchuk@fe.infn.it}

%==================================================================================================================================

\begin{abstract}
The rising phase of the 2023–24 outburst of the recently discovered bright transient black hole candidate Swift J1727.8-1613 was monitored 
by {\it Insight}-HXMT. We study the evolution of hard ($4$–$150$ keV) and soft ($2$–$4$ keV) band photon count rates, the hardness ratio (HR), 
QPO frequencies, and spectral features using daily observations from the HXMT/LE, ME, and HE instruments between August 25 and October 5, 2023. 
The QPO frequency is found to be strongly correlated with the soft-band X-ray count rates, and spectral photon indices.  
In contrast, a strong anti-correlation is observed between HR and QPO frequency, as well as between HR and photon index.  
Based on the evolution of the temporal and spectral properties, the rising phase of the outburst is subdivided into six parts. 
The evolution of the QPOs in parts 1–5 is fitted with the propagating oscillatory shock (POS) solution to understand the nature of the 
evolution from a physical perspective. An inward-propagating shock with weakening strength (except in part 4) is observed during the period of our study. 
The probable mass of the source is estimated to be $13.5 \pm 1.9~M_\odot$ using the QPO frequency ($\nu$)–photon index ($\Gamma$) scaling method.
\end{abstract}


\section{Introduction}

Stellar-mass black holes, especially transient black hole candidates (BHCs), are fascinating objects in X-rays as they exhibit rapid
evolution of timing and spectral properties during their active (i.e. outbursting) phases. Various temporal and spectral features,
along with their evolution, are generally observable during the outbursts of transient BHCs. These timing and spectral properties,
including quasi-periodic oscillations (QPOs), jets, and outflows, are found to be strongly correlated with each other 
\citep[see][]{RM06, Belloni05, Nandi12, D15, D21, Jana17}.
During a classical or type-I outburst of a transient BHC, four distinct spectral states are typically observed: hard (HS),
hard-intermediate (HIMS), soft-intermediate (SIMS), and soft (SS). These states form a hysteresis loop in the following sequence:
HS $\rightarrow$ HIMS $\rightarrow$ SIMS $\rightarrow$ SS $\rightarrow$ SIMS $\rightarrow$ HIMS $\rightarrow$ HS. However, in a
`failed' or type-II outburst, softer states (and sometimes even intermediate states) are found to be missing 
\citep[see][and references therein]{D17}. 

Although there is ongoing debate about what triggers an outburst in a transient BHC, it is generally believed that an outburst is
initiated by a sudden enhancement of viscosity at the outer edge of the disk \citep{Ebisawa96}. Recently, \citet{C19}
proposed that matter supplied by the companion accumulates at the pile-up radius ($X_p$), located at a far distance between the black hole
and the Lagrange point L1. During the quiescence phase, the accumulation of a large amount of matter at $X_p$ increases viscosity and
instability. When viscosity crosses a critical threshold, it triggers the onset of a new outburst.
In a type-I outburst, all accumulated matter at $X_p$ is cleared, whereas in a type-II outburst, matter is only partially cleared,
leaving behind residual material that is eventually expelled during the next type-I outburst along with freshly accumulated matter
\citep{C19,Bhowmick21,Chatterjee22}.
Recently, \citet{Russell18} developed a X-ray Binary New Early Warning System (XB-NEWS) for the possible outburst of a transient BHC 
by monitoring its increasing optical activity in the quiescence phase.
%The relaton between quiescence and outbursting phases and physcial explanation for the occurance of the classical and failed outbursts 
%are quite successfully been explained with this model for three well known recurring triansient BHCs such as H 1743-322 
%(Chakrabarti et al. 2019), GX 339-4 (Bhowmick et al. 2021), 4U 1630-472 (Chatterjee et al. 2022).

Low-frequency ($0.01$–$30$ Hz) quasi-periodic oscillations (LFQPOs) are commonly observed during the hard and intermediate spectral states of 
transient black hole candidates (BHCs). QPOs appear as peaks in the Fourier-transformed power-density spectra (PDS) of light curves, characterized 
by narrow noise components, and arise due to rapid variability in X-ray intensities. They provide crucial insights into the dynamics of accretion 
flows around black holes. Based on their properties (centroid frequency, Q-value, rms amplitude, noise, lag, etc.), LFQPOs are classified into three 
types: A, B, and C \citep{Casella05}. Generally, the primary frequency of type-C QPOs evolves monotonically during the HS and HIMS of both the 
rising and declining phases of an outburst. In contrast, type-B and type-A QPOs are sporadically observed in the SIMS.

Several models have been proposed to explain the origin of LFQPOs. The Lense-Thirring precession model \citep{Stella99} suggests that LFQPOs 
arise due to the precession of a tilted inner accretion flow caused by frame-dragging effects near the black hole. The precessing hot flow model 
\citep{Ingram11} proposes that a truncated disk, along with an inner hot flow, undergoes slow, global precession, modulating X-ray emissions. 
The accretion-ejection instability model \citep{Tagger99, Rodriguez02} suggests that LFQPOs result from magnetohydrodynamic instabilities 
in magnetized accretion disks. \citet{Titarchuk98} associated QPOs with viscous magneto-acoustic resonance oscillations in the transition layer 
surrounding the Compton cloud. \citet{Trudolyubov99} attributed LFQPOs to perturbations within a Keplerian disk, while \citet{Titarchuk00} 
explained them as global disk oscillations or oscillations of a warped disk.
According to the shock oscillation model (SOM) developed by Chakrabarti and collaborators \citep{MSC96,RCM97}, LFQPOs originate from shock 
oscillations in the accretion flow. In the Two-Component Advective Flow (TCAF) solution \citep{CT95}, a hot Comptonizing region, known as 
`CENBOL', naturally forms in the post-shock region. In the SOM framework, shock oscillations occur due to the heating and cooling effects 
within the CENBOL. The model suggests that sharp type-C QPOs arise from resonance oscillations of the shock, while type-B QPOs occur either 
due to the non-satisfaction of the Rankine-Hugoniot condition (necessary for stable shock formation) or due to a weakly resonating Compton 
cloud. Broader type-A QPOs are attributed to weak oscillations of the shockless centrifugal barrier \citep[see,][]{C15}. To explain evolution 
of QPO frequencies during rising and declining phases of transient BHCs a time varying form of the SOM namely propagating oscillatory 
shock (POS) model was introduced by Chakrabarti and his collaborators in 2005 \citep[see,][]{C05,C08}.

In astronomical sources, particularly in compact objects, mass is a crucial intrinsic parameter. Accurately determining the mass of the central 
black hole in a binary system is essential for understanding accretion-ejection processes. However, direct dynamical measurements of black hole 
mass are sometimes not possible due to the absence of detectable binary companion information (e.g., if the companion is too faint or non-detectable). 
In such cases, alternative methods are employed, including: QPO frequency ($\nu$)–photon index ($\Gamma$) correlation method \citep{ST07,ST09};
High-frequency QPO (HFQPO)–spin ($a$) correlation method \citep{Motta14}; Inverse scaling relation with observed HFQPOs \citep{RM06}; 
QPO frequency evolution method \citep{Iyer15, Molla16}; TCAF model-based spectral fitting method \citep{Molla16, Molla17}.
The spin and inverse scaling methods are not universally applicable, as HFQPOs are rare phenomena observed in only a limited number of BHCs. 
However, the $\nu$–$\Gamma$ correlation method has been widely used to estimate the masses of stellar-mass black holes, active galactic 
nuclei (AGNs), and neutron star systems. The TCAF model-based spectral fitting method has also been applied to estimate the probable masses 
of many transient BHCs.

Observationally, the spectral index increases as black holes transit from hard to intermediate to soft states. During the rising phase of transient 
BHCs, as the spectral state evolves from HS to HIMS, QPO frequencies exhibit a monotonic increase. Conversely, during the declining phase, as the 
source transits from HIMS to HS, QPO frequencies show a monotonic decrease. According to the TCAF model, during the rising phase of an outburst, 
an increasing rate of Keplerian matter (the source of thermal photons) pushes the shock inward, causing the Compton cloud to shrink. The enhanced 
supply of Keplerian matter increases cooling, reducing the proportion of hard photons and making the spectrum softer, thereby increasing the 
spectral index. Since QPO frequency follows an inverse relation with the shock location ($X_s$), i.e., $\nu \sim X_s^{-3/2}$, this results in a 
monotonic rise in QPO frequency. During the declining phase, the shock moves outward due to a rapidly decreasing supply of Keplerian matter, leading 
to a relative increase in the dominance of hard photons. As a result, QPO frequencies decrease while the spectrum hardens. In the SIMS, both spectral 
index and observed QPO frequencies are generally found within narrow ranges \citep[see][]{Nandi12}. \citet{Titarchuk98} proposed that the spectral 
index is a fundamental property of the corona, implying a direct correlation between the spectral index and QPO frequency. They also suggested that 
QPO frequency scales inversely with black hole mass, following the relation $\nu \sim M^{-1}$. Observationally, stellar-mass black holes 
($M_{\rm BH} \sim 10M_\odot$) exhibit LFQPOs in the $0.1$–$30$ Hz range and HFQPOs in the kHz range, while supermassive black holes 
($M_{\rm BH} \sim 10^6$-$10^9M_\odot$) exhibit LFQPOs in the mHz-$\mu$Hz range.

The Galactic transient BHC Swift J1727.8-1613 was discovered by Swift/BAT as a Gamma-Ray Burst candidate (GRB 230824A) on August 24, 2023 
\citep{Kennea23}. This bright outburst (maximum flux $\sim 7.6$ Crab) lasted for approximately nine months and was extensively studied across 
multiple wavelengths. The detection of its companion and radial velocity curve through spectro-photometric studies led to the dynamical confirmation 
of the source as a black hole by \citet{MataSanchez25}. They estimated an orbital period of $P_{\rm orb} = 10.81 \pm 0.001$~h and a lower mass limit 
of the black hole as $3.12 \pm 0.1~M_\odot$, including the mass function, $f(M_1) = 2.77 \pm 0.09~M_\odot$.
\citet{Peng24} estimated the spin ($a \sim 0.98$) and inclination angle ($i \sim 40^\circ$) of the source. 
However, \citet{D24} classified Swift J1727.8-1613 as a high-inclination BHC ($i \sim 85^\circ$), which aligns with the detection of soft 
time lags. The energy-averaged polarization degree ($4.1\% \pm 0.2\%$) and polarization angle ($2.2^\circ \pm 1.3^\circ$) were measured by 
\citep{Veledina23}. Polarization studies were also conducted by \citet{Podgorny24} and \citet{Ingram24}. LFQPOs have been detected in both 
soft and hard X-ray bands \citep{D24, Mereminskiy24, Nandi24, Zhu24}.

In this study, we analyze the evolution of the spectral and temporal properties during the rising phase of the 2023-24 outburst of 
Swift J1727.8-1613 using Insight-HXMT data. By studying the $\nu$–$\Gamma$ correlation using the scaling method, we estimate the mass of 
the source. Additionally, the study of QPO evolution using the propagating oscillatory shock (POS) model allows us to understand the evolution 
of the Compton cloud. The paper is organized as follows: \S 2 provides a summary of the POS model 
and $\nu$–$\Gamma$ correlation method. The \S 3 describes observations, data reduction, and analysis procedures. In \S 4, we present the results, 
while \S 5 discusses our findings and presents the conclusions.

\section{Models for Data Analysis}

\subsection{Propagating Oscillatory Shock (POS) Model}

\citet{C90} showed that, due to the satisfaction of the Rankine-Hugoniot conditions, unique standing shocks may form in the presence 
of a strong centrifugal barrier, where infalling energy may be dissipated, leading to the formation of jets and outflows. At the shock location 
($X_s$), matter that was flowing supersonically becomes subsonic and piles up in the post-shock region. This post-shock region, namely CENBOL, acts 
as the hot corona, up-scattering soft photons via inverse Comptonization. It may also oscillate if its cooling timescale roughly matches the infall 
time in this region \citep{MSC96}. Due to the non-satisfaction of the Rankine-Hugoniot conditions, the shock may also oscillate \citep{RCM97}. 
\citet{C15} observationally showed that resonance shock oscillation occurs only when the ratio between the Compton cooling time and the 
infall time falls within approximately 50\% of unity, i.e., between 0.5 and 1.5.

We observe variations in hard X-ray fluxes as the shock oscillates, i.e., as the size of the hot Compton cloud changes. During the oscillation phase, 
when the shock moves inward (due to an increase in the cooling rate with the rise in thermal seed photons from the Keplerian disk), the size of the 
Compton cloud shrinks. This reduces the interception of thermal photons as well as the number of emitted hard photons from Compton cloud or CENBOL. 
Similarly, when the shock moves outward (due to a faster decrease in the Keplerian disk accretion rate relative to the sub-Keplerian halo rate), the size 
of the Compton cloud increases, allowing more thermal photons to interact. This, in turn, increases the number of up-scattered hard photons from CENBOL. 
Overall, variations in the number of hard photons during the shock oscillation phases are reflected in the light curve, showing QPO-like features in 
the power density spectrum (PDS).

As the supply from the Keplerian disk and the sub-Keplerian halo varies during the outbursting phases, we generally observe the evolution of QPOs 
during both the rising and declining phases of an outburst. During the rising phase, the oscillating shock is found to move inward as the Keplerian 
disk accretion rate increases over time. Since the QPO frequency ($\nu_{QPO}$) is found to be inversely proportional to the infall timescale 
($t_{\text{infall}} \sim R X_s (X_s - 1)^{1/2}$, where $R(=\rho_+/\rho_-)$ is the shock compression ratio, i.e., the ratio between post-shock and 
pre-shock densities and $X_s$ is the shock location), we observe a monotonically increasing QPO frequency during the rising phase of an outburst. 
Similarly, during the declining phase of the outburst, an opposite trend is generally observed, as the shock is found to recede.

In \citet{C05,C08}, a modified time-varying shock oscillation model, i.e., the propagating oscillatory shock (POS) model, was 
proposed to study the evolution of monotonically evolving QPO frequencies during the rising and declining phases of the 2005 outburst of the 
Galactic transient BHC GRO J1655-40. According to this model, the frequency of the QPO ($\nu_{QPO}$) depends on the location ($X_s$) and 
strength ($\beta_s = 1/R$) of the shock, as well as on the mass of the black hole.

Thus, the instantaneous QPO frequency $\nu_{QPO}$ is given by
$$
\nu_{QPO} = \frac{\nu_{s0}}{t_{infall}}= \frac{c^3}{2GM_{BH}[R X_s (X_s-1)^{1/2}]}, 
\eqno{(1)}
$$
where $\nu_{s0} = c/r_s = c^3/2GM_{BH} = 10^5/(M_{BH}/M_{\odot})$ Hz is the inverse of the light crossing time of a black hole of mass $M_{BH}$ 
 and $c$ is the velocity of light, $r_s = 2GM_{BH}/c^2$ is the Schwarzschild radius. 
%The simplified form the above equation is
%$$
%\nu_{QPO} =\frac{C}{[R X_s (X_s-1)^{1/2}]},
%\eqno{(2)}
%$$
%
%where $\nu_{s0}$ is a constant $= 10^5/(M_{BH}/M_{\odot}){\rm s}^{-1}$.

In the drifting or evolving shock scenario, $X_s = X_s(t)$ is time-dependent and can be expressed as
$$
X_s(t) = X_{s0} \pm \frac{v(t) t}{r_s},
\eqno{(2)}
$$
where `$-$' sign is used for inward and '+' sign is used for receding shock motion. The shock velocity $v(t)$ may be 
accelerating, decelerating or constant and can be written as 
$$
v(t) = v_0 \pm v_a t, 
\eqno{(3)}
$$
where $v_0$ is the initial shock velocity and $v_a$ is the velocity acceleration (`+')/deceleration (`-') term. 

Generally during the rising phase of an outburst, the shock strength becomes weaker as the $R$ decreases. On the day 
of the highest evolving QPO, it becomes even weaker, approaching $R \sim 1$. The value of $R$ generally follows the equation 
$$
\beta_s = \frac{1}{R} = \frac{1}{R_0} + \alpha t^2,
\eqno{(4)}
$$
where $R_0$ is the initial compression ratio, $\alpha$ is the controlling factor of the reduction of the shock strength over time.

\subsection{QPO Frequency ($\nu$) - Photon Index ($\Gamma$) Correlation Scaling}

It is well established that the power-law photon index has a direct correlation with the properties of the Compton cloud or corona 
in black holes. The emitted hard X-ray fluxes are also found to be correlated with the size of the corona. This, in turn, affects the 
frequency of the observed QPOs. Titarchuk \& Fiorito (2004) introduced a model to study correlations between the observed QPO frequency 
($\nu$) and the model-fitted spectral photon index ($\Gamma$). Using a known source as a reference, this method is employed to estimate 
the masses of various compact objects, including stellar-mass BHs, AGNs, neutron stars, etc. \citep[see][]{ST07, ST09}. 
In this model, the central source is considered to be surrounded by a ``Compton cloud" along with a transition layer between this Compton 
cloud and the Keplerian disk. According to this model, the origin of the QPO frequencies is explained as the magneto-acoustic resonance 
oscillation frequency of the bounded transition layer \citep{Titarchuk02}, while the spectral photon index is attributed to changes 
in the size of the Compton cloud.

When studying the correlation between the QPO frequency and the photon index, it is generally observed that, before saturating at a certain value, 
the photon index initially exhibits an increasing trend with the QPOs. Thus, there are two primary phases: a trending phase and a constant phase, 
separated by the transition frequency ($\nu_{tr}$). 

The empirical relation given by \citet{ST07} is
$$
\Gamma(\nu) = A - D~B~ln\left[exp\left(\frac{\nu_{tr} - \nu}{D}\right) + 1\right],
\eqno(5)
$$
where $A$ represents the saturation level of the photon index, $B$ scales with the mass of the BH ($M_{BH}$) and the transition/threshold 
frequency above which the photon index saturates. The parameter $D$ controls the transition, i.e., how quickly the saturation is reached.

In \citet{ST09}, a more generalized version of this function is provided, making it applicable for estimating the mass 
and distance of an unknown compact object from the correlations of $\nu-\Gamma$, or $\nu-N_{bmc}$ (where $N_{bmc}$ is the bulk motion 
Comptonization model-fitted spectral normalization):
$$
f(x) = A - D~B~ln\left[exp\left(\frac{1- (x/x_{tr})^\beta}{D}\right) + 1\right],
\eqno(6)
$$
where $x$ is $\nu$ and $f(x)$ is either $\Gamma$ or $N_{bmc}$. 

Using Eqs. (5) and (6), one can fit the variation of the QPOs with $\Gamma$ or $N_{bmc}$ and extract a set of parameters: $A$, $B$, $D$, 
$\nu_{tr}$ or $x_{tr}$, $\beta$ (only from Eq. 6).

To estimate mass of an unknown source, $\nu-\Gamma$ variations are fitted using Eq. (6) for both unknown and known reference sources. 
According to \citet{ST09}, the scaling relation for the mass estimation is given by 
$$
M_U = \left(\frac{\nu_{trU}}{\nu_{trR}}\right) \times M_R,
\eqno(7)
$$
where suffixes $U$, $R$ mark unknown and known reference sources respectively. The ratio between transition frequencies, i.e., 
$s_\nu = \frac{\nu_{trU}}{\nu_{trR}}$ acts as a scaling factor to estimate the mass of the unknown source.

\begin{figure*}%[!h]
  \centering
    \includegraphics[angle=0,width=8.8cm,keepaspectratio=true]{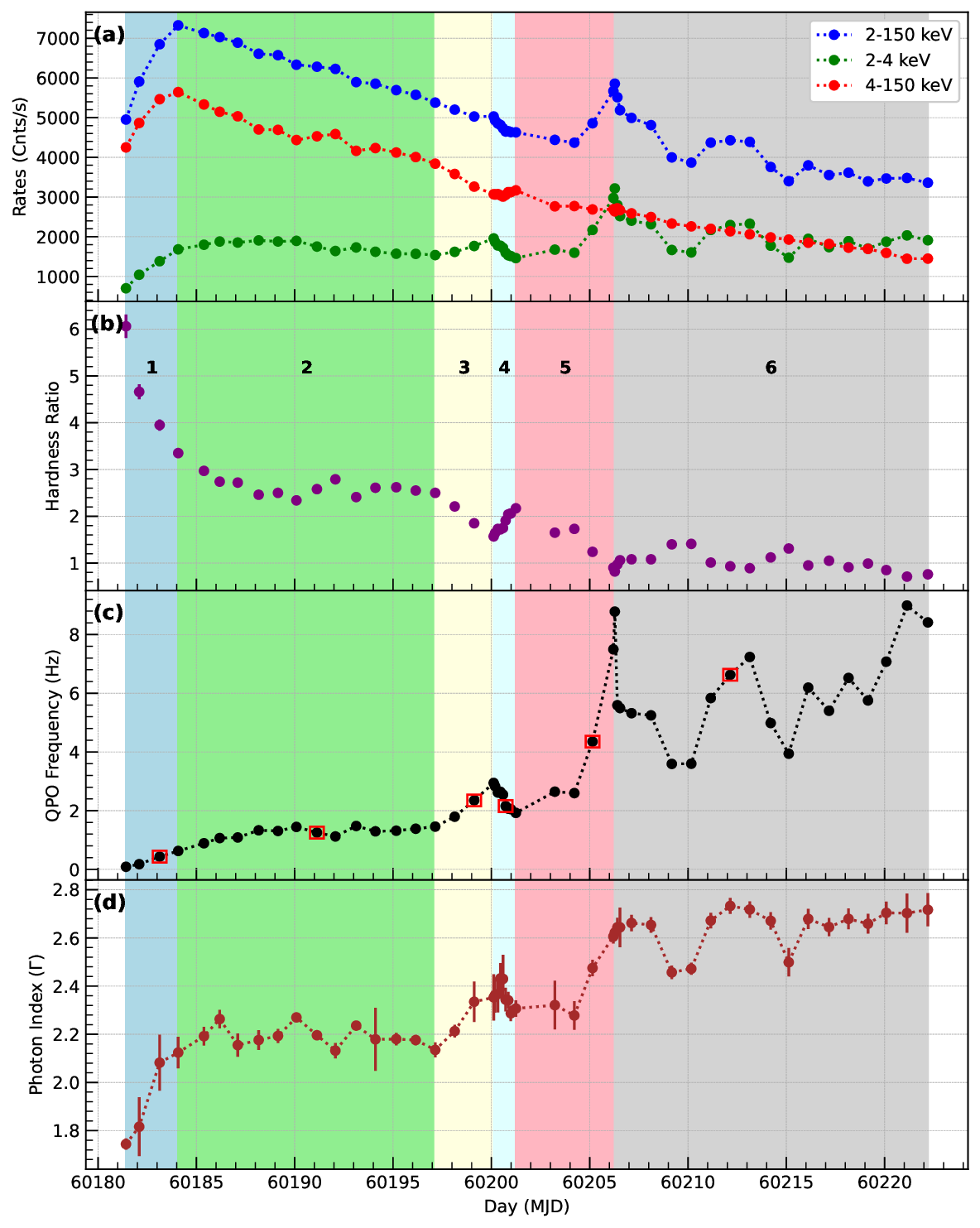}\hskip 0.2cm
    \includegraphics[angle=0,width=8.8cm,keepaspectratio=true]{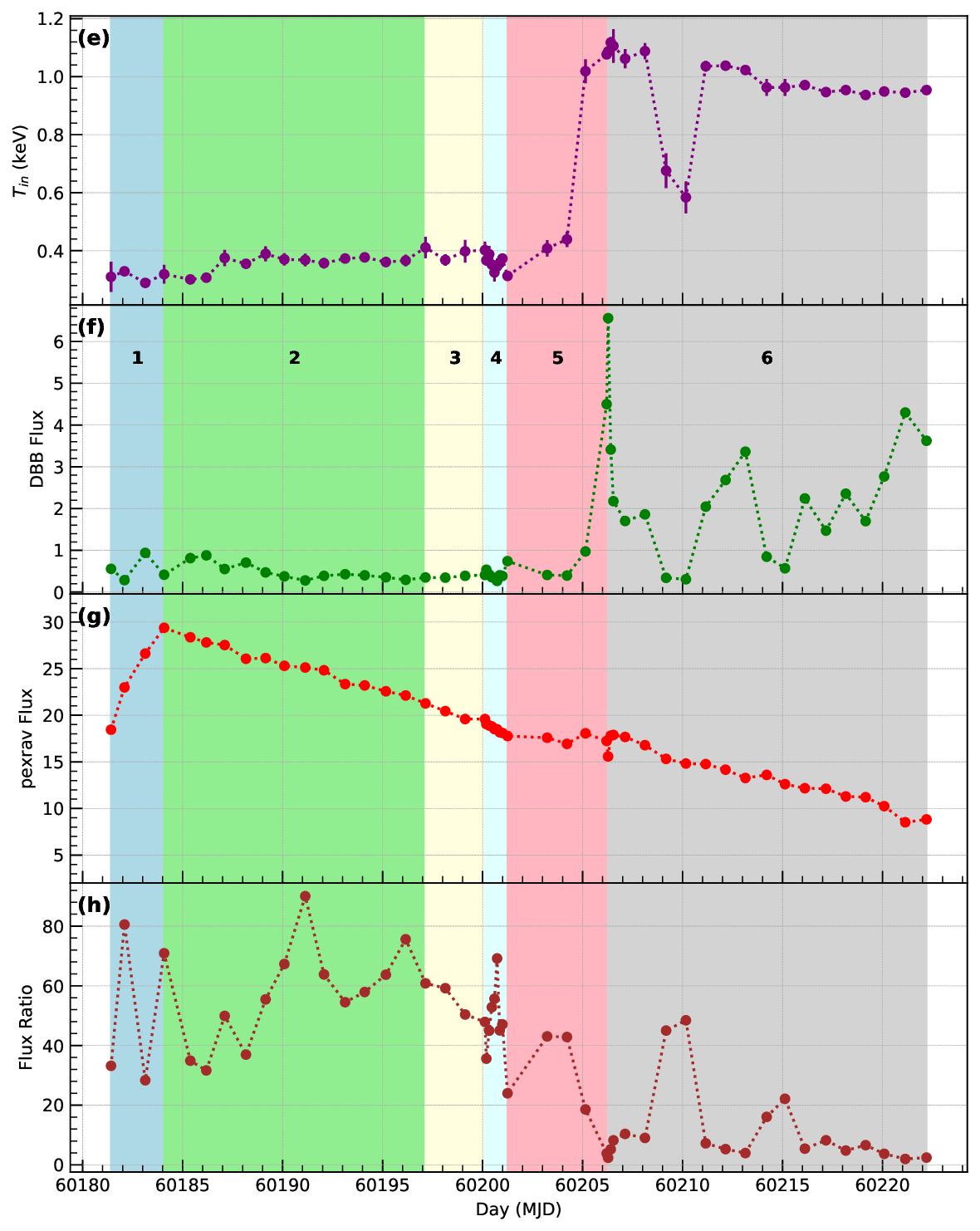}
    \caption{Variation of (a) {\it Insight}-HXMT count rates in the soft X-ray (SXR; $2$--$4$~keV), hard X-ray (HXR; $4$--$150$~keV), and 
	total (TXR; $2$--$150$~keV) energy bands, and (b) the hardness ratio (HR; i.e., the ratio between HXR and SXR rates) is shown.  
        (c) The evolution of the frequency of the observed QPOs is presented in the second lower-left panel.  
        Variation of the {\fontfamily{qcr}\selectfont constant$\otimes$tbabs(diskbb+pexrav) model-fitted parameters: (d) pexrav photon 
	index ($\Gamma$), (e) diskbb temperature ($T_{\rm in}$), and  (f)--(h) model component fluxes and their ratio are shown in 
	the subsequent panels.} Background shades subdivide the evolution of the QPO frequencies into six parts:  
	1: monotonically rising, 2: roughly constant, 3: monotonically rising, 4: monotonically decreasing, 5: rapidly rising, and 
	6: sporadically increasing or decreasing. 
	Furthermore, based on the temporal and spectral variations, parts 1--6 are reclassified as HS(ris), HIMS(ris), HIMS(ris), HIMS(dec), 
	HIMS(ris), and SIMS(ris), respectively. 
	The red-boxed points in panel (c) indicate the observations for which detailed PDS and spectral fits are presented in 
	Figures~\ref{fig_pds} and \ref{fig_spec} respectively.}
\label{fig_lc_spec}
\end{figure*}

\section{Observation and Data Analysis}

We study archival data of $51$ observations across all three energy band payloads (LE, ME, HE) of the {\it Insight}-HXMT \citep{Zhang14, Zhang20} 
starting from its first day of observation. One observation per day is used, except on 2023 Sep 13, where eight observations are taken. 
The dataset spans from 2023 Aug 25 (MJD=60181.42) to 2023 Oct 05  (MJD=60222.20). 

To study properties of QPOs, count rates, and hardness-ratios (HRs), we generated $1$~s and $0.01$~s time binned light curves in four 
different energy bands: $2-4$~keV of LE, $4-10$~keV of LE, $10-30$~keV of ME, and $30-150$~keV of HE. For the spectral analysis, 
we used data from all three instruments. The data extraction for generating light curves and spectral files was performed using 
{\tt hpipeline} pipeline within the HEASoft software package (v.6.34) from HeaSARC.

To study the variation of the count rates and hardness ratios, we use $1$ s time-binned light curves. For power density spectra (PDS), we use
$0.01$ s time-binned light curves in the ME energy band. To determine the parameters (centroid frequency, FWHM, power) of the QPOs, the PDS 
are fitted with the Lorentzian model in the XRONOS package of HEASoft.

The combined HEXTE/LE ($2-10$~keV), ME ($10-35$~keV), and HE ($33-150$~keV) data, covering a broad energy range of $2-150$~keV, are fitted with  
{\fontfamily{qcr}\selectfont constant$\otimes$tbabs(diskbb+pexrav)}. A variable hydrogen column density ($N_H$) in the range of 
$(0.5-4) \times 10^{22}$~${\rm atoms~cm^{-2}}$ is used for the absorption model {\tt tbabs}. Since a strong reflection feature with a cutoff 
power-law at higher energies is present throughout the entire rising phase, we use the {\tt pexrav} model \citep{Magdziarz95} to fit the spectra. 
The best fits are obtained based on the reduced $\chi^2$ ($\chi^2_\nu$) statistics ($\sim 1$).  

\newpage
\section{Results}

Detailed temporal and spectral properties during the rising phase of the outburst of Swift~J1727.8$-$1613 are studied using broadband 
{\it Insight}-HXMT data. Based on the evolution of spectral and temporal features, we divided the rising phase of the outburst into six parts 
with date range: $i)$ MJD = 60181.42--60184.07, $ii)$ MJD = 60184.07--60197.15, $iii)$ MJD = 60197.15--60200.12, $iv)$ MJD = 60200.12--60201.25,  
$v)$ MJD = 60201.25--60206.28, and  $vi)$ MJD = 60206.28--60222.20. Parts 1--3 and 4--6 are further classified as stage~I and stage~II, respectively. 
The detailed characteristics of these stages, along with the estimation of the mass based on combined temporal and spectral analysis, are discussed 
in the following sub-sections.

\subsection{Outburst Profile}

We used $1$~s time-binned light curves from all three energy band instruments of {\it Insight}-HXMT to study the outburst profile during the 
rising phase (MJD=60181.42 - 60222.20) of Swift J1727.8-1613. In Fig.~\ref{fig_lc_spec}(a), we show the variation in the count rates in Soft X-Ray 
(SXR; $2-4$~keV), Hard X-Ray (HXR; $4-150$~keV), and Total X-Ray (TXR; $2-150$~keV) energy bands. 

All three bands showed an initial rise. The evolution of these three rates can be classified into two stages. In stage-I (up to 2023 Sep. 13; 
MJD=60200.12, i.e, parts 1-3), TXR and HXR followed a similar trend, whereas in stage-II (from 2023 Sep. 13 to Oct. 05; MJD=60200.12-60222.20, 
i.e., parts 4-6), TXR and SXR showed roughly similar variability. 

The total and HXR band rates peaked on the 4th day (MJD=60184.07), whereas SXR showed the maximum rate on the 6th day (MJD=60186.19). 
There is a roughly $2.12$~day gap between the peaks of the HXR and SXR band rates. According to \citet{Jana16}, this might indicate an important 
accretion flow parameter, namely the viscous timescale, which represents how high-viscosity Keplerian disk matter flows during the outburst.

\begin{figure}%[!h]
  \centering
    \includegraphics[angle=0,width=9.0cm,keepaspectratio=true]{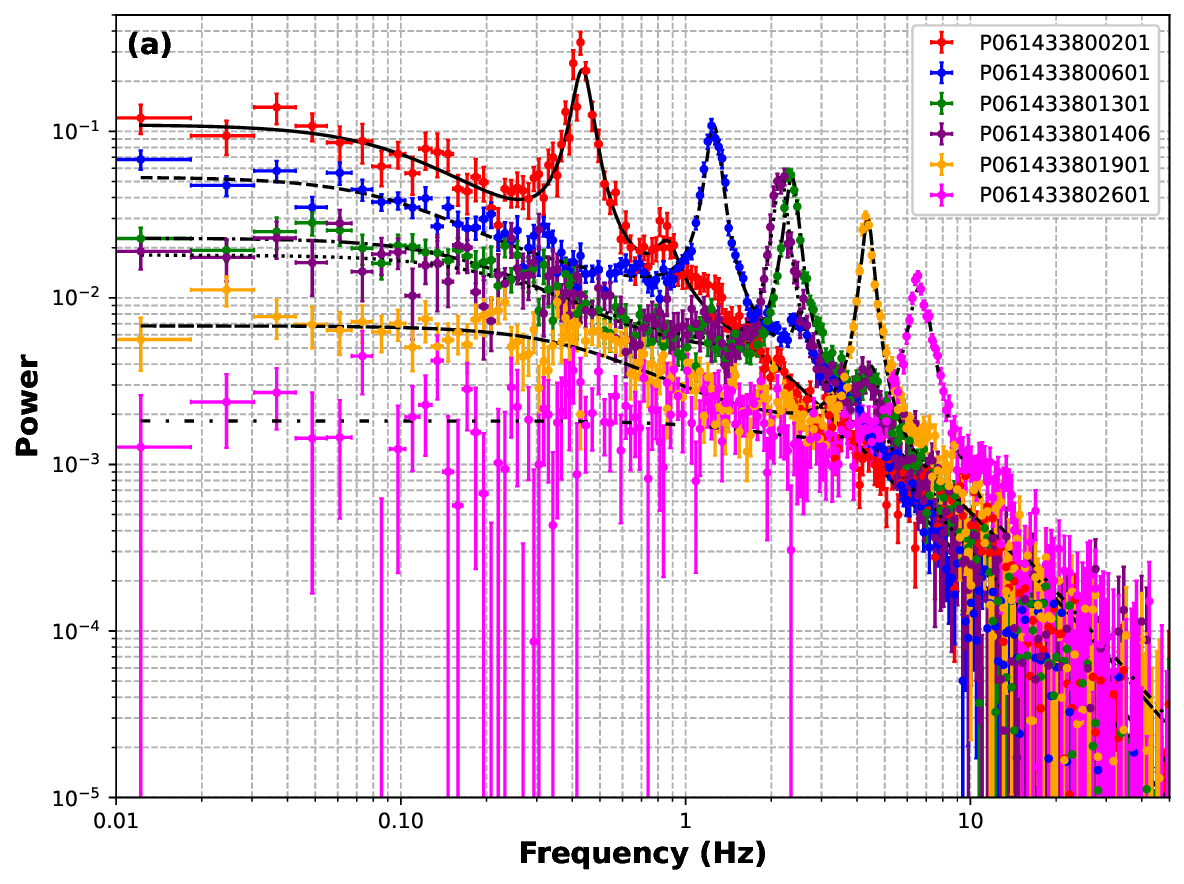}\vskip 0.02cm
    \includegraphics[angle=0,width=9.0cm,keepaspectratio=true]{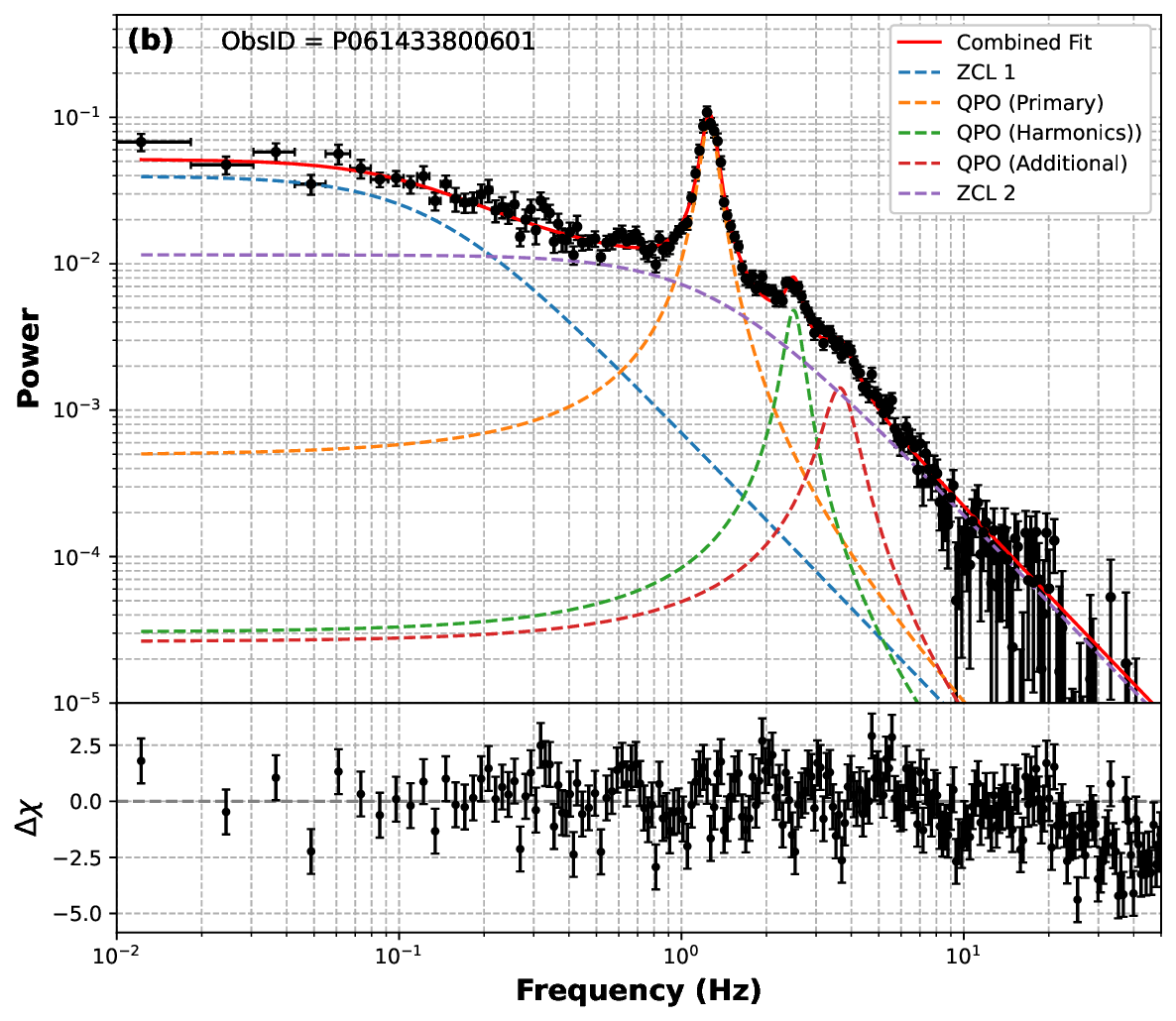}
	\caption{(a) Fourier-transformed power density spectrum of the $0.01$~s time-binned $10$–$30$~keV HXMT/ME light curves
	from six observations selected from six different parts of the rising phase (red boxed points of Fig.~\ref{fig_lc_spec}c). 
	(b) Bottom panel shows detailed Lorentzian model fitted total (solid red curve) and components (two zero-centric and three 
	QPO-centric; dashed cruves) for ObsID = P061433800601.} 
	%Avg_MJD:  60183.1313, 60191.1311, 60199.1275,   60200.7263,  60205.1457, 60212.1504
\label{fig_pds}
\end{figure}

\subsection{Hardness-Ratio (HR)} 

The variation of the hardness ratio (HR) or X-ray color-color ratio is shown in Fig.~\ref{fig_lc_spec}(b). We used the ratio between the broad-band 
$4-150$~keV energy HXR count rate (using the combined light curves of all three HXMT instruments) with $2-4$~keV band HXMT/LE SXR count rate as HR. 
A sharp decrease in HR ($6.06-2.97$) is observed within $\sim 4$ days between 2023 Aug 25-29 (MJD = 60181.42-60185.38). Then, HR remains nearly 
constant around $2.5$ for the next $\sim 12$ days until 2023 Sep 10 (MJD = 60197.15). Subsequently, HR shows a decreasing trend ($2.50-1.57$) 
for the next $\sim 3$ days until 2023 Sep 13 (MJD = 60200.12).

After that, a sharp rise in HR ($1.57-2.17$) is observed over the next $\sim 1.3$ days until 2023 Sep 14 (MJD = 60201.25). Then, a decreasing 
trend in HR ($2.17-0.82$) is found over $\sim 5$ days until 2023 Sep 19 (MJD = 60206.28). After that, HR is observed to fluctuate within a wide 
range of $0.71-1.41$ until the last day of our observation (MJD = 60222.20).

\subsection{Low Frequency QPOs}

A strong signature of sharp type-C QPOs is observed throughout the entire observation period. The evolution of the frequency of the primary QPOs 
is shown in Fig.~\ref{fig_lc_spec}(c). During the six parts of the rising phase of the outburst, the QPO frequency exhibits the following 
trends: $i)$ monotonically rising from $0.09$ to $0.63$~Hz, $ii)$ slowly increasing from $0.63$ to $1.46$~Hz, $iii)$ monotonically rising in the 
range $1.46$--$2.94$~Hz, $iv)$ monotonically decreasing in the range $2.94$--$1.92$~Hz, $v)$ rapidly rising from $1.92$ to $8.78$~Hz, and  
$vi)$ sporadically increasing or decreasing between $3.94$ and $8.99$~Hz.  

In Fig.~\ref{fig_pds}, we show the model-fitted six PDS, selected from each part of the rising phase of the outburst. Detailed model components 
and the variation of $\Delta \chi$ are also shown for part~2 ObsID: P061433800601. The Lorentzian model-fitted parameters, $Q$-value (defined as 
FWHM/$\nu_{\rm QPO}$), and the fractional rms percentage are listed in Appendix Table~\ref{table_qpo_spec}.

Parts 1-5 are defined as the evolving phase, during which the frequency of the observed QPOs evolves from $0.09$~Hz (on 2023 Aug 25; MJD = 60181.42)
to $8.78$~Hz (on 2023 Sep 19; MJD = 60206.28). We further studied this evolving phase of the QPOs using the propagating oscillatory shock (POS) solution
to understand the nature of the evolution from a physical perspective. Part 6 is defined as a non-evolving or sporadic phase since no clear trend
is observed.

\subsection{Spectral Result}
\begin{figure}%[!h]
\vskip -0.2cm
  \centering
    \includegraphics[angle=0,width=9.2cm,keepaspectratio=true]{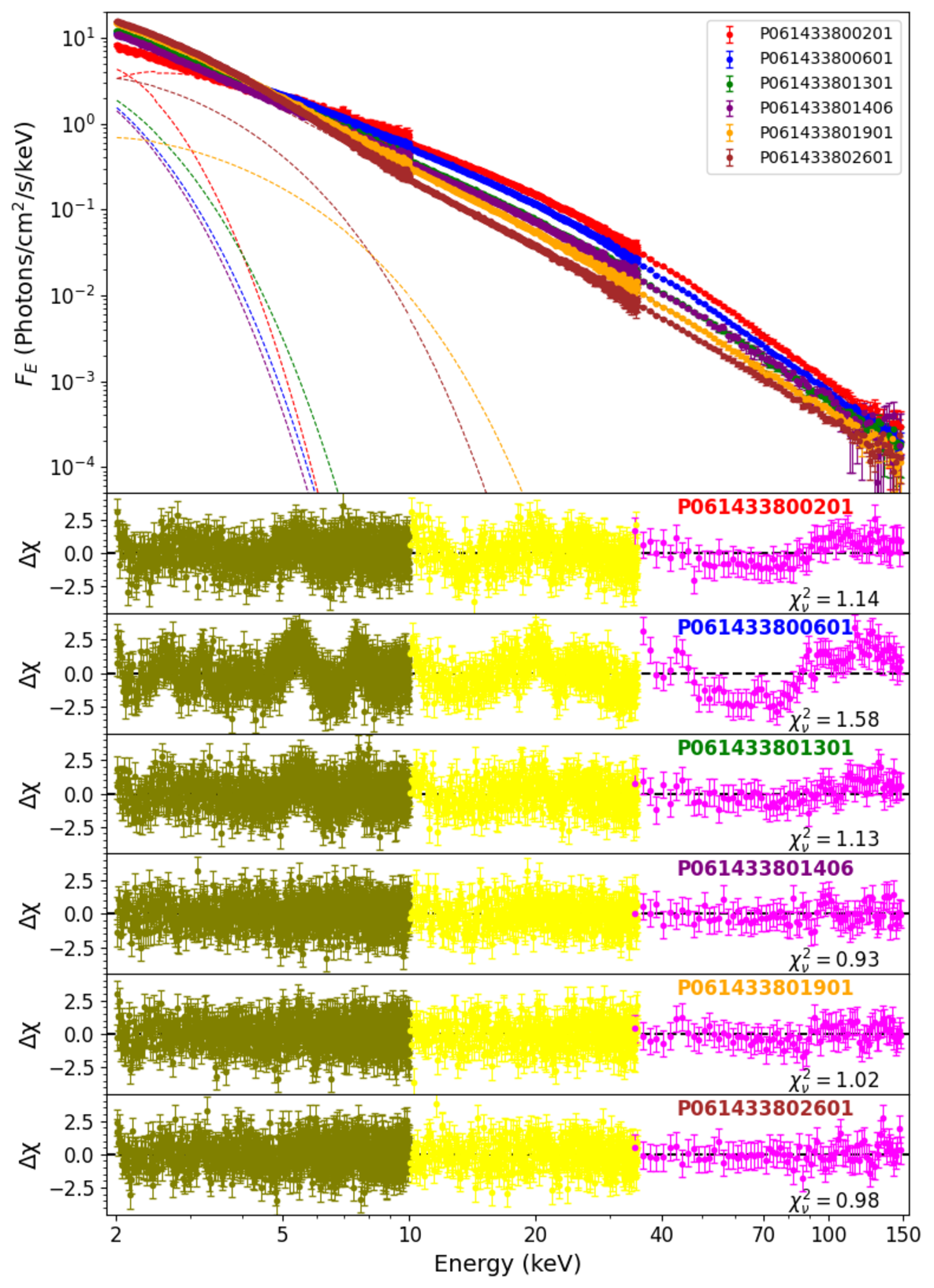}
	%\vskip -0.5cm
	\caption{Broadband spectral fits using \textit{Insight}-HXMT LE (2–10 keV), ME (10–35 keV), and HE (33–150 keV) data with the 
	{\fontfamily{qcr}\selectfont tbabs(diskbb+pexrav)} model for six representative spectra, each selected from a different part of the rising 
	phase of the 2023-24 outburst of Swift~J1727.8$-$1613.
	}
\label{fig_spec}
\end{figure}

Broadband spectra using LE ($2$--$10$~keV), ME ($10$--$35$~keV), and HE ($33$--$150$~keV) data from {\it Insight}-HXMT during the entire 
rising phase of Swift~J1727.8$-$1613 are studied using the {\fontfamily{qcr}\selectfont constant$\otimes$tbabs(diskbb+pexrav)} model. 
In Fig.~\ref{fig_spec}, we show the model-fitted spectra from six observations, each selected from a different part of the rising 
phase of the outburst. The variation of the model-fitted \texttt{diskbb} (DBB) temperature ($T_{\rm in}$) and \texttt{pexrav} photon index 
($\Gamma$) parameters is shown in Fig.~\ref{fig_lc_spec}(d--e). In the later panels (f--h) of Fig.~\ref{fig_lc_spec}, we show the variation 
of the model component flux contributions within the studied $2$--$150$~keV band, along with the flux ratio between the \texttt{pexrav} and 
\texttt{diskbb} components.

In part~1, similar to the trends observed in count rates of different energy bands, both $\Gamma$ and the \texttt{pexrav} model flux are found 
to increase. The photon index $\Gamma$ rises from $1.74$ to $2.12$. This phase can be defined as the rising hard state, i.e., HS(ris). On the transition 
day (MJD = 60184.07) between parts~1 and~2, the TXR, HXR, and \texttt{pexrav} fluxes reach their peak. After this, the \texttt{pexrav} flux begins 
to decrease which continued till the end of analysis period. A similar trend is also found in the HXR. 
In part~2, $\Gamma$ remains roughly constant at $\sim 2.2$, while in part~3, it increases from $2.14$ to $2.36$. During parts~1--3, $T_{\rm in}$ 
and the \texttt{diskbb} flux remain nearly constant at lower levels, and the flux ratio fluctuates at higher values. Spectrally, parts~2 and~3 are 
identified as the rising hard-intermediate state, i.e., HIMS(ris).

Subsequently, during the short duration of part~4 ($\sim 1.1$~days), the source exhibits behavior reminiscent of the declining phase of a 
canonical outburst in classical or type-I transient black hole candidates \citep[see, e.g.,][]{Nandi12,D17}. A rapid decrease in $\Gamma$, $T_{\rm in}$, 
and both model component fluxes is observed. Meanwhile, the flux ratio increases, indicating that the \texttt{DBB} flux decreases more rapidly than 
the \texttt{pexrav} flux. This suggests that the source is in the declining hard-intermediate state, i.e., HIMS(dec). In this phase, a decreasing trend 
in QPO frequency, HXR, and TXR, along with an increasing trend in SXR and HR, is also observed.

\begin{figure*}%[!h]
\vskip -0.2cm
  \centering
    \includegraphics[angle=0,width=16.5cm,keepaspectratio=true]{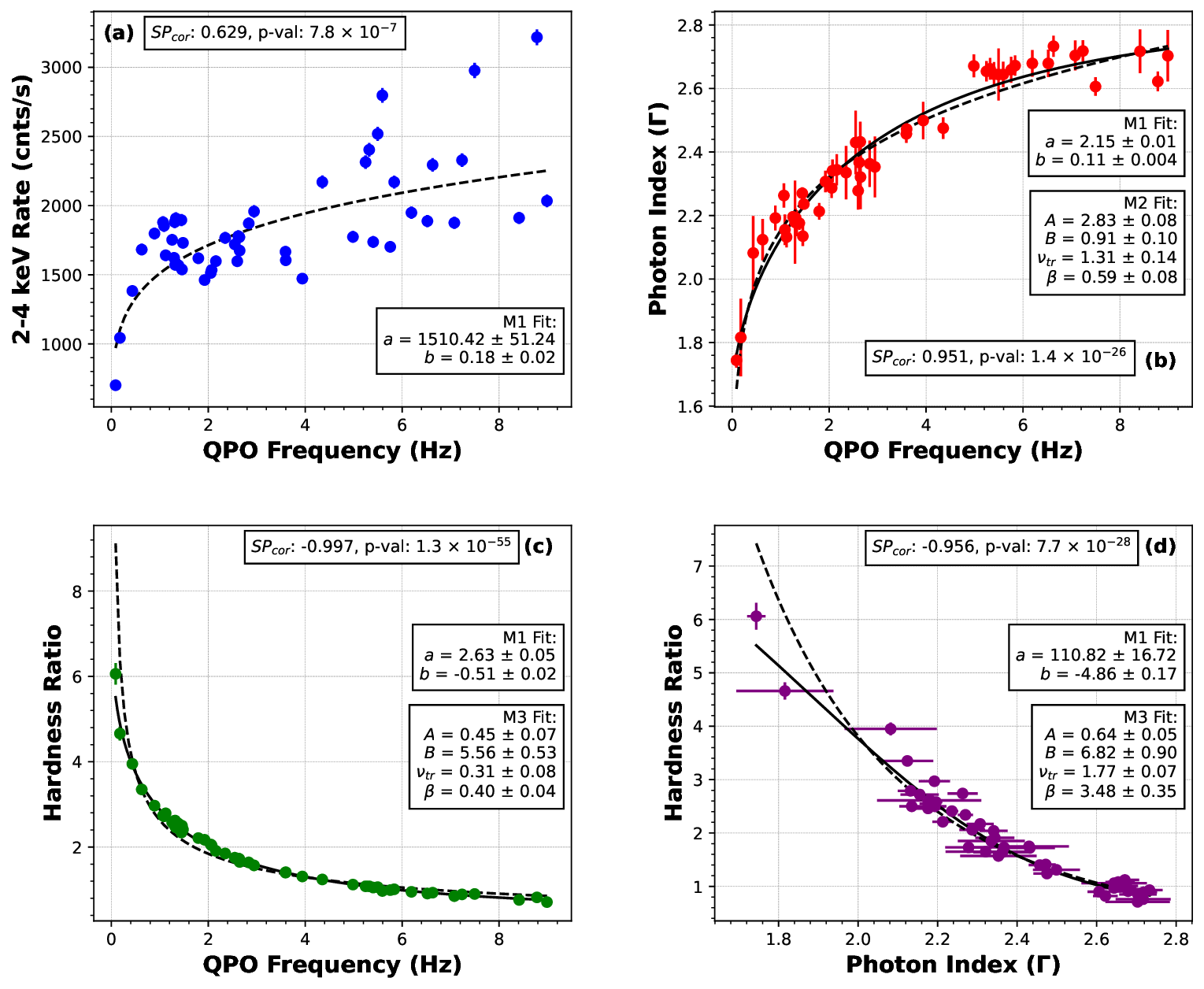}
	%\vskip -0.5cm
	\caption{The variation of QPO frequency ($\nu_{QPO}$) with (a) Soft X-ray (SXR; $2-4$~keV) count rate, (b) photon index ($\Gamma$), and 
	(b) Hardness ratio (HR), defined as the ratio of Hard X-ray (HXR; $2-150$~keV) to SXR rates is shown. In (d), the variation of HR 
	with $\Gamma$ is presented. %Pearson and Spearman rank correlation coefficients, i.e., $P_{\text{cor}}$ and $SR_{\text{cor}}$, along with 
	%their p-values, are provided in the insets. The results indicate strong positive correlations of $\nu_{QPO}$ with SXR and $\Gamma$, whereas 
	%HR exhibits strong negative correlations with both $\nu_{QPO}$ and $\Gamma$. 
	The power-law model ($y = ax^b$), referred to as M1, is used to obtain a crude fit to the correlations (shown as dashed curves). 
	We further fitted the $\nu_{\text{QPO}}$–$\Gamma$ correlation using the scaling relation of Eq.~(6), referred to as M2, and the HR correlations 
	with $\nu_{\text{QPO}}$ and $\Gamma$ using Eq.~(8), referred to as M3 (shown as solid curves). In both M2 and M3 models, the parameter 
	$D$ was frozen at unity.
	}
\label{fig_cor}
\end{figure*}

In part~5, the source again shows characteristics similar to those of the canonical rising harder spectral states. We observe a rapid increase 
in $\Gamma$, $T_{\rm in}$, and the \texttt{diskbb} flux. The flux ratio decreases and reaches a minimum on the transition day between parts~5 and~6.  
Spectrally, this part of the outburst can again be identified as HIMS(ris). In part~6, except for a late-stage plateau, $\Gamma$, $T_{\rm in}$, and 
the flux ratio vary sporadically. While the \texttt{pexrav} flux shows a monotonic decrease, the \texttt{diskbb} flux displays irregular variations.  
This phase of the outburst is classified as the soft-intermediate state, i.e., SIMS(ris).

\subsection{Correlation of QPO frequency with SXR and Photon Index}

It is quite evident from Fig.~\ref{fig_lc_spec} that the evolution of the QPO frequency follows a trend roughly similar to that of the soft X-ray rate (SXR) 
in the $2-4$~keV HEXTE/LE band. The soft X-rays primarily originate from the Keplerian disk, which cools the Compton cloud (or corona) during the 
process of producing higher-energy photons via inverse-Compton scattering. Thus, the SXR has a direct effect on the size of the Compton cloud, which is 
also responsible for the origin of low-frequency QPOs (LFQPOs) according to the shock oscillation model (SOM; see Eq.~1). To statistically confirm the 
dependence of SXR on the QPO frequency ($\nu_{\rm QPO}$), we studied the correlation between them using the Spearman rank correlation method 
(see Fig.~\ref{fig_cor}a). Indeed it indicates that there exists a strong positive correlation. The correlation coefficient is fund to be as 
$SR_{\text{cor}} = 0.629$ with a p-value of $7.8 \times 10^{-7}$.  

The variation of the $pexrav$ model-fitted photon index ($\Gamma$) exhibits a trend roughly similar to that of $\nu_{\text{QPO}}$, except during 
the late rising phase (see Fig. ~\ref{fig_cor}d). The Spearman rank correlation analysis was performed on the variation of $\nu_{\text{QPO}}$ vs. $\Gamma$ 
(see Fig.~\ref{fig_cor}b). The correlation coefficient is found as $SR_{\text{cor}} = 0.951$ with a p-value of $1.4 \times 10^{-26}$, which also suggests 
presence of a strong positive correlation between $\nu_{\text{QPO}}$ and $\Gamma$. 

Furthermore, both correlations were first studied using a simple power-law model ($y = ax^b$), referred to as the M1 model, which yields a 
positive index ($b \sim 0.11$--$0.18$) in both cases. The $\nu_{\text{QPO}}$--$\Gamma$ correlation is found to be tighter and was further fitted 
with the empirical scaling formula given in Eq.~(6), referred to as the M2 model. In Fig.~\ref{fig_cor}(a--b), the dashed curves represent the M1 
model fits, while in Fig.~\ref{fig_cor}(b), the solid curve shows the M2 model fit with fixed D=1. Note that due to large scattered 
variation in the $\nu_{\text{QPO}}$--SXR plot, an acceptable fit could not be obtained using the M2 model.

\subsection{Anti-correlation of HR with QPO frequency and Photon Index}
The hardness ratio (HR) exhibits a trend opposite to that of the QPO frequency (see Fig.~\ref{fig_lc_spec}). From a simple observation, this feature 
appears to be more prominent during the evolving phase of the QPO frequency (parts 1-5). To confirm this statistically, we analyzed the Spearman 
rank correlation. It indicates a strong negative (anti-)correlation between the QPO frequency and the hardness ratio (HR), with a correlation coefficient 
of $SR_{\text{cor}} = -0.997$ and a p-value of $1.3 \times 10^{-55}$. As in the previous correlation studies shown in Fig.~\ref{fig_lc_spec}(a--b), 
we modeled this correlation using the power-law form, which yields a negative index of $b = -0.51 \pm 0.05$.

Since the trend between the QPO frequency and HR follows an approximately exponential decay, with a slight indication of late saturation at higher 
QPO values. To study the relationship between $\nu_{\text{QPO}}$ and HR, we fitted it with the modified the scaling equation (6) by changing 
the sign between two factors on the right-hand side of the correlation formula (defined as M3 model). The modified equation is given by:  
$$
f(x) = A + D~B~\ln\left[\exp\left(\frac{1- (x/x_{\text{tr}})^\beta}{D}\right) + 1\right],
\eqno(8)
$$  

where $x$ represents $\nu_{\text{QPO}}$, and $f(x)$ corresponds to the calculated HR.  

A strong anti-correlation trend between $\nu_{\text{QPO}}$ and HR is well fitted by the above Eq. (8), yielding a reduced $\chi^2$ value of $1.01$. 
The best-fit parameters with fixed D$=1$ are found to be: $A = 0.45 \pm 0.07$, $B = 5.57 \pm 0.53$, $\nu_{\text{tr}} = 0.31 \pm 0.08$, and 
$\beta = 0.40 \pm 0.04$. In Fig.~\ref{fig_cor}(c), $\nu_{\text{QPO}}$–HR data points (green) along with the model-fitted curves (dashed curve 
for M1 and solid curve for M3) are shown.  

While studying the correlation between HR and the photon index ($\Gamma$) with the Spearman rank correlation method, we found a strong anti-correlation 
as correlation coefficient is obtained as $SR_{\text{cor}} = -0.956$ with a p-value of $7.7 \times 10^{-28}$. The variation of $\Gamma$ vs. HR is also 
well fitted with the both M1 and M3 models (see Fig.~\ref{fig_cor}d). The best-fitted M3 model parameters with D=1 are found to be: $A = 0.64 \pm 0.05$, 
$B = 6.82 \pm 0.90$, $\Gamma_{\text{tr}} = 1.77 \pm 0.07$, and $\beta = 3.48 \pm 0.35$. A steeper negatve powerlaw index ($b= -4.86\pm0.17$) 
is found while fitting correlation with the M1 model. In Fig.~\ref{fig_cor}(d), $\Gamma$–HR data points (purple) along with the model-fitted curves 
(dashed curve for M1 and solid curve for M3) are shown.  

\begin{figure}%[!t]
\vskip -0.2cm
  \centering
    \includegraphics[angle=0,width=9.2cm,keepaspectratio=true]{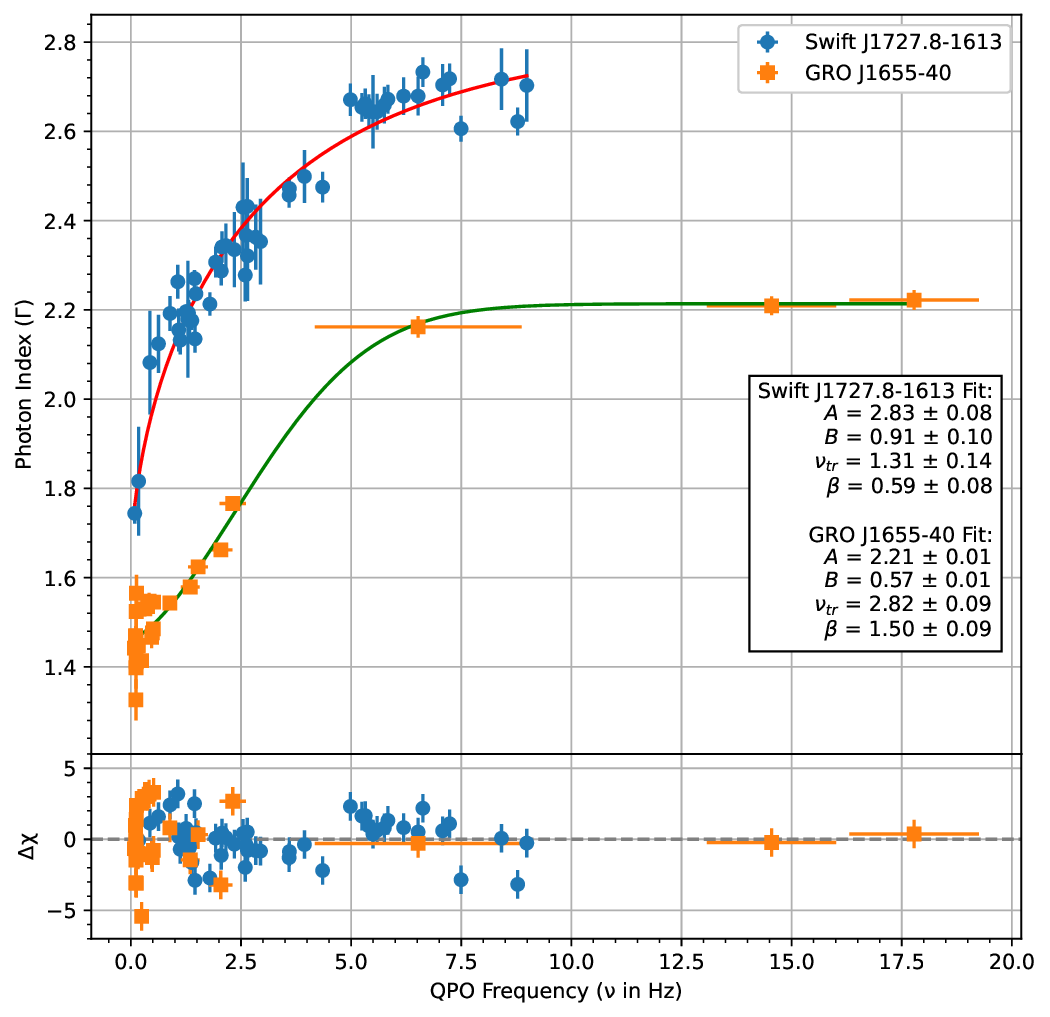}
	\caption{Photon index ($\Gamma$) vs. QPO frequency ($\nu_{QPO}$) correlation scaling for the estimation of the mass of BHC Swift J1727.8-1613 
	using GRO J1655-40 as the known ($M_{BH}=6.3\pm0.5~M_\odot$) reference source. Using model (Eq. 6 with parameter D fixed at 1) fitted 
	$\nu_{tr}$ values of GRO J1655-40 and Swift J1727.8-1613, mass of the unknown BHC Swift J1727.8-1613 is estimated to be $13.5\pm1.9~M_\odot$. 
	For GRO J1655-40, $\nu_{QPO}$ and $\Gamma$ values are adopted from the rising phase data of 2005 outburst (see, Chakrabarti et al. 2008, 
	Debnath et al. 2008). 
         }
\label{fig_nuga}
\end{figure}

The QPO evolution during the entire rising phases of both the unknown source Swift J1727.8-1613 (current 2023-24 outburst) and the reference source 

\subsection{Mass Estimation From $\nu_{QPO}$ - $\Gamma$ Scaling Method}

According to Section 5.1 of \citet{SS73}, all characteristic dynamical timescales in accreting matter onto a compact object are determined 
by the mass of the object. The low-frequency QPO is also found to be inversely proportional to the mass of the black hole (BH). Therefore, one can
scale the mass of the BH using the observed QPO frequency. Various methods for mass estimation based on low- and high-frequency QPOs have been
proposed. \citet{ST09} introduced a more generalized method for mass scaling of an unknown source using a reference source
with a known mass. The details of this method are already discussed in \S2.2.

Here, we used the rising phase data of the QPO frequency and photon index (adopted from \citealt{C08} and \citealt{D08} from the
2005 outburst of GRO J1655-40 as the reference source's data. GRO J1655-40 is chosen as the reference since it exhibits a similar variation of 
$\nu-\Gamma$ as the target source Swift J1727.8-1613, and its mass is well known from dynamical methods (Greene et al. 2001). The Swift J1727.8-1613 
photon index is obtained from spectral fitting in a broad energy range of $2-150$~keV {\it Insight}-HXMT data. For spectral fit, we use the combined 
model: {\fontfamily{qcr}\selectfont constant$\otimes$tbabs(diskbb+pexrav)}. The best-fitted spectra using the above combined model provide photon index 
values, which are used here to study the $\nu-\Gamma$ correlation as well as mass scaling using \citet{ST09} method.

GRO J1655-40 (2005 outburst) is fitted with the $\nu-\Gamma$ correlation formula as defined in Eq. 6 (see, Fig.~\ref{fig_nuga}). While fitting, 
we use a fixed value of $D (=1)$. The set of model-fitted parameters: $A$, $B$, $\nu_{tr}$, and $\beta$ are obtained for both sources. For 
Swift J1727.8-1613, the model-fitted values are as follows: $A: 2.83 \pm 0.08$, $B: 0.91 \pm 0.10$, $\nu_{tr}: 1.31 \pm 0.14$, $\beta: 0.59 \pm 0.08$. 
For the reference source GRO J1655-40, the corresponding values are: $A: 2.21 \pm 0.01$, $B: 0.57 \pm 0.01$, $\nu_{tr}: 2.82 \pm 0.09$, 
$\beta: 1.82 \pm 0.09$. Using the scaling formula in Eq. 7 and the mass of the reference source GRO J1655-40 as $6.3 \pm 0.5M_\odot$, we estimate 
the mass of the newly discovered Galactic transient black hole Swift J1727.8-1613 to be $13.5 \pm 1.9~M_\odot$. 
%This value is in good agreement with the mass obtained from the POS model fitted QPO evolution.

\begin{table}%[!h]
\addtolength{\tabcolsep}{-3.0pt} % to reduce gap between columns
\vskip -0.3cm
\centering
%\raggedright  % <-- left justify the table
%\caption{Primary QPO Information}
\caption{POS Model Fitted Initial Parameters}
\label{table_pos_para}
%\small
%\vskip -0.5cm
\renewcommand{\arraystretch}{0.9}  % to reduce gap between rows
\begin{tabular}{lccccccc} 
%\begin{tabular}{|l|cc||c|c|c|c|c|} 
 \hline
Parts& $M_{\rm BH}$ & $X_{s0}$ & $v_0$ & $v_a$ & $R_0$ & $\alpha$ & $v_{0,fitted}$ \\
\hline
1 & 13.5 & 773 & 14000 & -1900 &  4.0 & 0.0060 & 14314$^{\pm46}$\\
2 & 13.5 & 189 &   150 &    0.0 & 2.9 & 0.0020 & 173$^{\pm3}$\\
3 & 13.5 & 155 &   500 &    0.0 & 2.7 & 0.0095 & 480$^{\pm19}$\\
4 & 13.5 & 120 &  1500 &    0.0 & 2.0 & 0.0020 & 1495$^{\pm111}$\\
5 & 13.5 & 158 &    95 &    100 & 2.0 & 0.0195 & 97$^{\pm2}$\\
\hline
\end{tabular}
%\noindent{
\leftline{The units of fixed BH mass $M_{\rm BH}$ in $M_\odot$; shock location} 
\leftline{$X_{s0}$ in $r_s$, velocity $v_0$ in $cm~s^{-1}$, acceleration/deceleration} 
\leftline{$v_a$ in $cm~s^{-2}$, and strength controlling factor $\alpha$ in $s^{-2}$.}
\leftline{$v_{0,fitted}$ are model fitted inititial shock valocities in $cm~s^{-1}$.}
\end{table}

\begin{figure}%[!h]
\vskip -0.2cm
  \centering
    \includegraphics[angle=0,width=9.2cm,keepaspectratio=true]{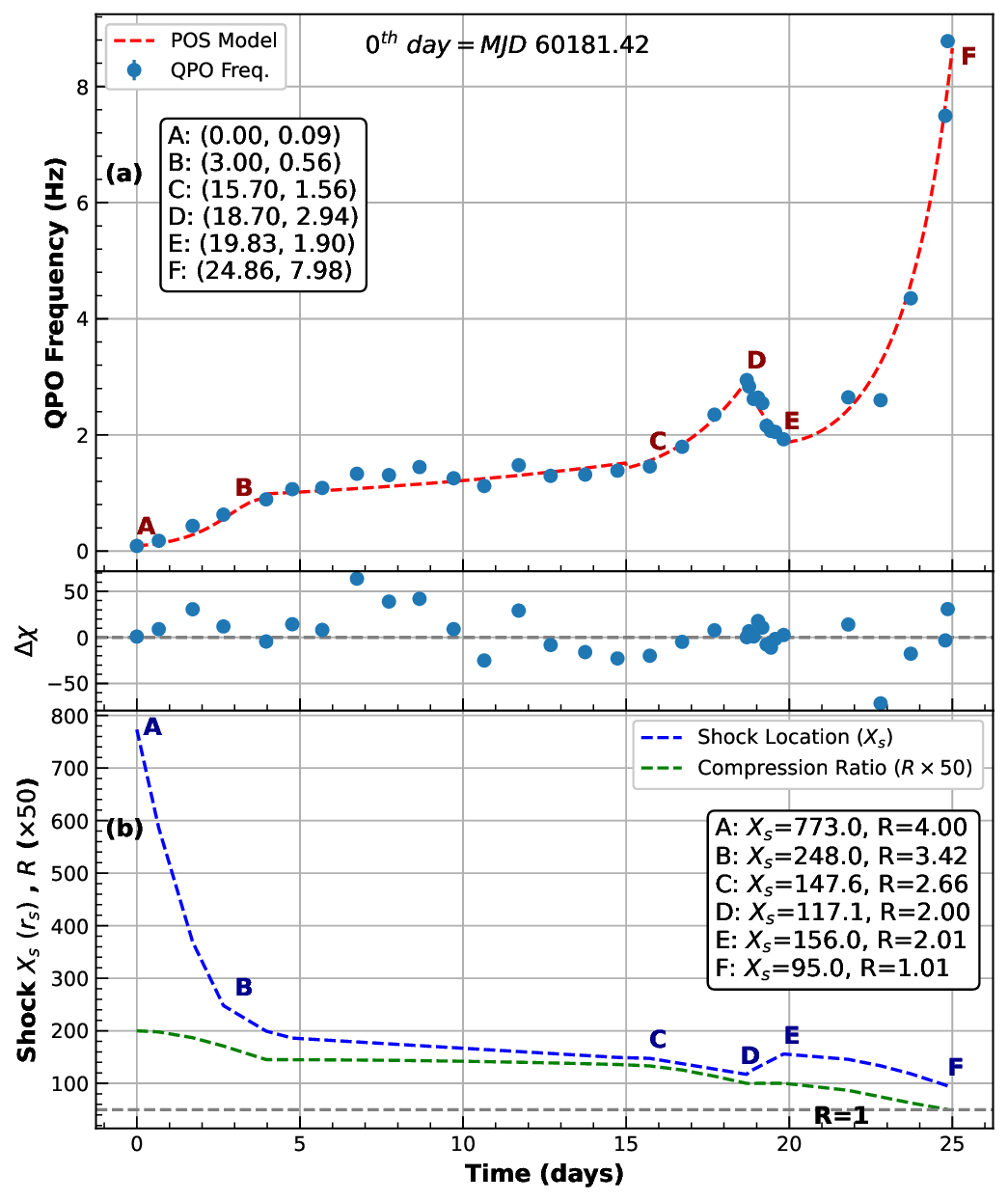}
	\caption{(a) Variations of the observed QPO frequencies with time (in day) from shaded parts (1-5) of Fig.~\ref{fig_lc_spec}, in the rising 
	phase of the outburst with the fitted POS model (dashed curve) are shown. (b) Variation of the shock location (in Schwarzschild 
	radius $r_s$) and shock compression ratio (R) are shown. The model fitted $\Delta \chi$ variation is shown in the middle panel. 
	Here, points A–F mark the start, end, or major transition phases of the evolution.
      		}
\label{fig_pos}
\end{figure}

\subsection{POS model fitted QPO evolution}

The propagating oscillatory shock (POS) model is the time-varying form of the shock oscillation model (SOM), which explains the origin of QPOs 
due to the oscillation of the shock, i.e., the outer radius of the hot Compton cloud or CENBOL. Thus, the oscillation of the shock corresponds 
to the oscillatory variation in the size of Compton cloud. During the rising phase of transient BHCs, the oscillating shock is found to move 
inward due to an increase in the rate of Keplerian disk matter. According to SOM, $\nu_{\text{QPO}} \sim X_s^{-3/2}$. 
As the shock moves inward, an increase in QPO frequencies is observed during the rising phase of an outburst. In the declining phase, the oscillating 
shock is found to move outward due to a rapid decrease in the rate of Keplerian disk matter. Consequently, in this phase, a monotonic decrease 
in QPO frequencies is observed over time.

This model has been used to study the evolution of QPOs during the rising and/or declining phases of outbursts in several sources, including 
GRO J1655-40 \citep{C05, C08}, GX 339-4 \citep{D10, Nandi12}, H 1743-322 \citep{D13}, and XTE J1550-564 \citep{C09}. The model fit provides 
the instantaneous location, strength, and velocity of the shock. \citet{Iyer15} and \citet{Molla16} used this model to estimate the masses of 
the BHCs IGR J17091-3624, and MAXI J1659-152 respectively. These studies motivated us to investigate the evolution of QPO frequencies 
during the rising phase of Swift J1727.8-1613 using the POS model.

The evolving QPOs in parts (1–5) of the rising phase of the outburst of Swift~J1727.8–1613 are fitted with the POS model (see Fig.~\ref{fig_pos}),
keeping the BH mass fixed at the $\nu$–$\Gamma$ scaling-derived value of $13.5~M_\odot$. The initial guesses for the other model-fitted parameters 
are listed in Table~\ref{table_pos_para}. Since acceptable fits could not be obtained when all model parameters were kept free, we refitted the 
evolution of the QPOs by allowing only the initial velocity ($v_0$) to vary. The best-fit values of the initial shock velocities ($v_{0,fitted}$) are 
provided in Table~\ref{table_pos_para}. Note that while fitting the rising phase QPO evolution, on the first day of the evolution (when QPOs are generally 
observed at frequencies $< 0.1$~Hz), the POS model assumes a large Compton cloud with a stronger shock strength ($R \sim 4$), whereas on the last day of 
the evolution, it assumes a smaller Compton cloud with the weakest shock strength ($R \sim 1$).

From the POS model fit, except in part 4, the shock is found to move inward with decreasing strength. Over $\sim 25$ days of the evolving phase, 
the shock location $X_s$ decreases from $\sim 773r_s$ to $95r_s$, accompanied by a change in the compression ratio from its stronger ($R=4$) to its 
weakest ($R=1$) possible value. In Fig.~\ref{fig_pos}, letters A-F denote the start, end, and transition days of different phases of QPO evolution. 
The shock is found to move inward rapidly in part 1, whereas in part 5, the shock weakens at a faster rate.

%From the best-fit POS model to the QPO frequency evolution, the mass of the BHC Swift~J1727.8$-$1613 is found to be $13.3 \pm 0.1~M_\odot$. 
%The confidence limits on the POS model–derived black hole mass are assessed using two methods: the covariance matrix and the profile likelihood approach. 
%Both methods yield broadly consistent estimates for $M_{\rm BH}$.
%%To confirm this estimated mass, we applied the scaling method as defined in \S2.2.

\section{Discussion and Concluding Remarks}

The discovery outburst of the Galactic transient BHC Swift~J1727.8-1613 was first detected on 2023 Aug 24 (MJD = 60180), and it continued for the 
next $\sim 10$ months. In this {\it paper}, we present a detailed study of the outburst profiles, hardness ratios, and low-frequency QPOs during 
the rising phase (2023 Aug 25 to Oct 05; MJD = 60181.42-60222.20) of the outburst using broadband X-ray data from the {\it Insight}-HXMT instruments 
in the $2-150$~keV energy range.

The daily variations in X-ray intensities in the soft ($2-4$~keV, SXR), hard ($4-150$~keV, HXR), and total ($2-150$~keV, TXR) energy bands show 
two distinct stages. In stage-I (up to 2023 Sep 13; MJD = 60200.12), TXR followed the HXR trend, while in stage-II (from 2023 Sep 13 to Oct 05; 
MJD = 60200.12-60222.20), TXR followed the SXR trend. From a physical perspective, we can infer that in stage-I, sub-Keplerian halo matter dominated, 
while in stage-II, Keplerian disk matter became dominant. Initially, both SXR and HXR rates increased and peaked on two different dates. The time gap 
between the HXR peak (MJD = 60184.07) and the SXR peak (MJD = 60186.19) allows us to estimate the viscous timescale of Swift J1727.8-1613 during its 
discovery outburst as $\sim 2.1$~days.
Except for the initial and late rise in stage-I, the SXR rate remains roughly constant, indicating a steady supply of Keplerian matter from the 
pile-up radius. Subsequently, in stage-II, SXR exhibits multiple peaks, suggesting an uneven supply of disk matter. 
At the beginning of stage-II, the SXR rate declines sharply, indicating a sudden decrease in viscosity. On the other hand, in the post peak 
dates, the HXR rate decreases monotonically, except for two instances: in the middle of stage-I and at the beginning of stage-II.

Based on the evolution of QPO frequencies, count rates, HRs, spectral parameters ($\Gamma$, $T_{\rm in}$), and two different types (thermal 
and nonthermal) model component fluxes, we classify the rising phase of the outburst into six parts. Parts (1-5) are termed the ``evolving phase", 
while part 6 is categorized as the ``non-evolving/sporadic phase". QPO frequencies increase during parts 1-3 and 5, whereas in part 4, they decrease 
over time. During part 4, we observe a sharp rise in HR due to the increasing HXR trend and the simultaneous decreasing trend of SXR. From the evolution 
of QPOs and HRs in parts 1-3 and 5, we infer that the oscillating shock responsible for generating the observed LFQPOs moves inward due to the increasing 
cooling rate, as indicated by the relative increase in SXR over time. This also suggests that the size of the hot Compton cloud, which serves as the 
reservoir of hot electrons that upscatter thermal photons from the Keplerian disk, reduces over time. However, in part 4, the shock moves outward, 
leading to an increasing Compton cloud size due to a reduced cooling rate, as evidenced by the increase in HXR and the decrease in SXR.

To understand the accretion flow dynamics of the source during different parts of the rising phase of the outburst, a joint \texttt{diskbb} 
and \texttt{pexrav} model-fitted spectral study has been performed using broadband {\it Insight}-HXMT data (see Fig.~\ref{fig_lc_spec}(d-h) and 
Appendix Table~\ref{table_qpo_spec}). Both the \texttt{pexrav} model-fitted photon index ($\Gamma$) and flux are found to increase rapidly during part~1, which 
is defined as HS(ris). A maximum in the \texttt{pexrav} model flux is observed on the transition day between parts~1 \& 2, after which the flux decreases 
monotonically. Such a local maximum in the sub-Keplerian halo accretion rate is generally observed on the HS$\rightleftharpoons$HIMS transition day during 
a canonical outburst of transient BHCs \citep{D18}. Based on this behavior and nature of the QPOs, the spectral states of both parts~2 \& 3 are defined 
as HIMS(ris). The \texttt{diskbb} temperature ($T_{\rm in}$) and flux are found to remain roughly constant at lower values during stage-I (i.e., parts~1--3). 
During this stage, the flux ratio (pexrav to \texttt{diskbb}) is observed to remain at higher values.
Subsequently, in the shorter interval of part~4, the source exhibited behavior characteristic of the declining phase of a canonical outburst. During this 
phase, not only the QPO frequency, SXR, and TXR decrease, but the spectral parameters ($\Gamma$, $T_{\rm in}$) and both model component fluxes also decline. 
Since the rate of decrease in \texttt{diskbb} flux is greater than that of \texttt{pexrav} flux, the flux ratio shows an increasing trend. This phase is 
classified as HIMS(dec). Next, part~5 is defined as HIMS(ris), as the source resumes a rising trend in HIMS behavior. A sharp increase is observed in 
$\Gamma$, $T_{\rm in}$, and \texttt{diskbb} flux. The sharp rise in thermal \texttt{diskbb} flux significantly reduces the size of the corona due to 
enhanced cooling, which is reflected in the sharp increase in the QPO frequency ($\nu_{\rm QPO}$). In the final phase (part~6), except for a late, 
roughly constant trend, sporadic increases and decreases are observed in $\Gamma$, $T_{\rm in}$, and \texttt{diskbb} flux. Consequently, the QPO frequency 
during this phase also shows a sporadic increasing or decreasing behavior. This phase is therefore defined as SIMS(ris).

Throughout the entire rising phase, the observed QPO frequencies evolve in a manner similar to that of the SXR and photon index, and inversely with the 
hardness ratio (HR) (see Fig.~\ref{fig_lc_spec}). To statistically confirm these trends, we performed correlation studies using the Spearman rank correlation 
method (see Fig.~\ref{fig_cor}). The analysis reveals a strong positive correlation between QPO frequency and both SXR and $\Gamma$, and a strong negative 
correlation between HR and both QPO frequency and $\Gamma$. The correlation and anti-correlation trends of QPO frequency, photon index, and HR values 
were further fitted using a simple power-law model (M1), as well as empirical relations defined in Eqs.~6 (M2) and 8 (M3). From the M1 fits, small positive 
indices ($b \sim 0.11$--$0.18$) are found for the correlations of $\nu_{\rm QPO}$ with SXR and $\Gamma$, while negative indices are obtained for the 
correlations of HR with both $\nu_{\rm QPO}$ and $\Gamma$. The M2 and M3 model fits of $\nu_{\rm QPO}$ with $\Gamma$ and HR clearly indicate initial 
rising trends followed by late saturation at `A' values after the transition frequencies $\nu_{\rm tr}$. This implies that as QPO frequency increases, 
$\Gamma$ and HR transition from a trending phase to a constant saturation phase. Although there is a slight indication, the M3 fit does not clearly 
exhibit a saturation trend in the correlation between $\Gamma$ and HR.

To estimate the mass of the source Swift~J1727.8$-$1613, we studied the $\nu$–$\Gamma$ correlation using the scaling method described in \S2.2.
Here, we use the 2005 rising phase data of the well-known transient BH GRO J1655-40 as the reference source to scale the mass of the unknown BHC 
Swift J1727.8-1613. The QPO frequency vs. photon index evolution of both sources is fitted with Eq. 6 to determine the scaling model parameters. 
Using the best-fitted transition frequencies ($\nu_{tr}$) of the two sources and the dynamically measured mass of the reference source, we estimate 
the mass of Swift J1727.8-1613 as $13.5\pm1.9~M_\odot$. By combining the POS model and $\nu-\Gamma$ scaling methods, we infer the probable mass of 
the newly discovered Galactic transient BHC Swift J1727.8-1613 as $13.5\pm1.9~M_\odot$. This result roughly agrees with \citet{D24} estimated source 
mass ($10.2\pm0.4~M_\odot$) from the spectral analysis with the TCAF model.

We further study the evolution of QPO frequencies during the evolving phase (parts 1–5) using the propagating oscillatory shock (POS) model. The 
instantaneous location and strength of the shock, responsible for generating the observed LFQPOs, are obtained from the model fit by keeping 
the black hole mass frozen at the value of $13.5~M_\odot$ as measured from the $\nu$–$\Gamma$ correlation scaling method.
During the entire evolving phase of $\sim 25$~days, the shock moves inward from $X_s = 773r_s$ to $95r_s$ while weakening in strength. The shock 
compression ratio $R$ (inverse of the strength $\beta_s$) decreases from its stronger value ($R=4$) to its weakest possible value ($R=1$). In part 4, 
which lasts $\sim 1.1$~days, the shock recedes from $117r_s$ to $156r_s$. Additionally, we observe that in part 1, the shock moves rapidly inward, 
whereas in part 5, the shock weakens at a faster rate. 
%The best-fitted POS model for the QPO frequency evolution also allows us to infer the probable mass of the BHC Swift J1727.8-1613 as 
%$13.3\pm0.1~M_\odot$, since the mass of the BH is a crucial parameter in the POS model (see \S2.1).

A brief summary of our findings in this {\it paper} is as follows:

\begin{enumerate}[i)]
    \item The rising phase of the outburst exhibits two distinctly different stages based on the variations in soft (SXR; 2-4 keV), hard
          (HXR; 4-150 keV), and total (TXR; 2-150 keV) X-ray rates. In stage-I, TXR correlates with HXR, while in stage-II, TXR correlates with SXR.
    \item The viscous timescale of the source is estimated to be $\sim 2.1$ days, based on the observed delay of approximately 2.12 days 
	  between the peaks of the HXR and SXR band rates.
    \item Strong signatures of type-C LFQPOs are observed throughout the rising phase. We further classify the rising phase of the outburst into 
	  six parts based on the evolution of $\nu_{\rm QPO}$, HR, count rates, and spectral features.
    \item Detailed spectral and timing analyses confirm the spectral states corresponding to these six parts of the outburst. We observe the 
	  spectral state evolution in the following sequence: HS(ris) $\rightarrow$ HIMS(ris) $\rightarrow$ HIMS(dec) $\rightarrow$ HIMS(ris) 
	  $\rightarrow$ SIMS(ris). Note that both parts~2 \& 3 correspond to HIMS(ris).
    \item A strong positive correlation of $\nu_{QPO}$ with SXR as well as the photon index is noticed.
    \item A strong negative or anti-correlation of HR with $\nu_{QPO}$ as well asp the photon index is observed.
    \item The POS model fit of $\nu_{\text{QPO}}$ during the evolving phase (parts~1–5) allows us to determine the instantaneous location, strength, 
	  and velocity of the shock, whose resonant oscillation is considered to be the origin of the observed strong type-C QPOs.
	  %It also provides the most probable mass of the BHC as $13.3\pm0.1~M_\odot$.
    \item The POS model-fitted QPO evolution suggests that during parts~1--3 and 5, the shock moves inward with increasing $\nu_{\rm QPO}$, 
	  whereas in part~4, a receding shock is inferred as $\nu_{\rm QPO}$ decreases. No clear trend in $\nu_{\rm QPO}$ is observed in part~6.
    \item The observation of a receding shock, accompanied by a monotonically decreasing trend in QPO frequency over a short period of 
	  $\sim 1.1$ days in part~4, indicates an unusual evolution — a signature of a declining phase occurring within the rising phase of the outburst.
    \item The $\nu-\Gamma$ scaling method enables us to estimate the mass of the newly discovered BHC Swift J1727.8-1613 as $13.5\pm1.9~M_\odot$,
          using GRO J1655-40 as the reference source.
\end{enumerate}

\section*{Acknowledgements}

We are thankful to the anonymous referee for his/her kind suggestions to improve the quality of the paper.
This work made use of archival data of {\it Insight}-HXMT, a mission satellite project of China National Space Administration
(CNSA) and Chinese Academy of Sciences (CAS).
D.D. acknowledge the visiting research grant of National Tsing Hua University, Taiwan (NSTC NSTC 113-2811-M-007-010). 
H.-K. C. is supported by NSTC of Taiwan under grant NSTC 113-2112-M-007-020.
S.K.N. acknowledges support from the visiting research grant 
of National Tsing Hua University.

%\appendix
%\renewcommand{\thetable}{A\arabic{table}} % For tables in appendix
%\setcounter{table}{0} % Reset counter so first appendix table is A1

\begin{table*}%[!h]
\renewcommand{\thetable}{A\arabic{table}} % Makes numbering A1, A2, etc.
\setcounter{table}{0} % Reset counter so first appendix table is A1
\addtolength{\tabcolsep}{-3.0pt} % to reduce gap between columns
\vskip -0.0cm
\centering
%\caption{Primary QPO Information}
\caption{QPO and Spectral Fitted Result}
\label{table_qpo_spec}
\small
%\scriptsize
\vskip -0.3cm
\renewcommand{\arraystretch}{0.9}  % to reduce gap between rows
\begin{tabular}{lcccc|ccc|ccccccc} 
%\begin{tabular}{|l|cc||c|c|c|c|c|} 
 \hline
Sr.& ObsID & UT      & MJD$_{Avg}$ & Exp & $\nu_{QPO}$ & Q & rms(\%) & N$_H$ & T$_{in}$ & $\Gamma$ & $rel_{\rm refl}$ &  $\chi^2_{\nu}$ & $F_{\rm DBB}$ & $F_{\rm pexr}$  \\
%No.&       & (mm/dd) &             & (s) & (Hz)        &   &          \\
	\multicolumn{5}{c|}{\textsc{Data Information}}& \multicolumn{3}{c|}{\textsc{QPO Result}}& \multicolumn{7}{c}{\textsc{Spectral Result}} \\
\hline
 1&X0101& 08/25& 60181.42& 565& 0.088$^{+0.002}_{-0.004}$&2.27$^{+0.56}_{-0.41}$& 15.2$^{+2.3}_{-1.8}$ & 3.33$^{\pm 0.19}$& 0.31$^{\pm 0.05}$& 1.74$^{\pm 0.02}$& 0.70$^{\pm 0.10}$ & 1.23 &0.56 &18.5 \\
 2&X0106& 08/26& 60182.09& 476& 0.177$^{+0.005}_{-0.006}$&3.79$^{+1.59}_{-0.81}$& 15.9$^{+4.9}_{-3.7}$ & 1.87$^{\pm 0.31}$& 0.33$^{\pm 0.02}$& 1.82$^{\pm 0.12}$& 0.93$^{\pm 0.12}$ & 1.06 &0.28 &23.0 \\
 3&X0201& 08/27& 60183.13& 516& 0.433$^{+0.005}_{-0.005}$&3.99$^{+0.68}_{-0.55}$& 18.8$^{+2.3}_{-1.9}$ & 3.53$^{\pm 0.32}$& 0.29$^{\pm 0.01}$& 2.08$^{\pm 0.12}$& 2.64$^{\pm 0.41}$ & 1.14 &0.94 &26.6 \\
 4&X0208& 08/28& 60184.07& 476& 0.627$^{+0.006}_{-0.007}$&8.64$^{+2.76}_{-1.99}$& 17.6$^{+4.4}_{-3.4}$ & 1.23$^{\pm 0.40}$& 0.32$^{\pm 0.03}$& 2.12$^{\pm 0.07}$& 5.25$^{\pm 0.63}$ & 1.12 &0.41 &29.4 \\
 5&X0301& 08/29& 60185.38& 557& 0.890$^{+0.004}_{-0.004}$&6.30$^{+0.53}_{-0.47}$& 19.7$^{+1.3}_{-1.2}$ & 2.10$^{\pm 0.15}$& 0.30$^{\pm 0.01}$& 2.19$^{\pm 0.04}$& 2.90$^{\pm 0.47}$ & 1.30 &0.81 &28.4 \\
 6&X0307& 08/30& 60186.19& 451& 1.066$^{+0.004}_{-0.004}$&7.11$^{+0.59}_{-0.53}$& 19.5$^{+1.2}_{-1.2}$ & 2.31$^{\pm 0.16}$& 0.32$^{\pm 0.01}$& 2.26$^{\pm 0.04}$& 3.08$^{\pm 0.34}$ & 1.25 &0.88 &27.8 \\
 7&X0314& 08/31& 60187.10& 400& 1.087$^{+0.006}_{-0.006}$&7.37$^{+1.10}_{-0.88}$& 18.6$^{+2.0}_{-1.8}$ & 0.89$^{\pm 0.17}$& 0.38$^{\pm 0.03}$& 2.16$^{\pm 0.05}$& 3.71$^{\pm 0.65}$ & 1.44 &0.55 &27.5 \\
 8&X0408& 09/01& 60188.17& 450& 1.331$^{+0.004}_{-0.005}$&7.21$^{+0.54}_{-0.49}$& 18.8$^{+1.1}_{-1.0}$ & 1.21$^{\pm 0.12}$& 0.36$^{\pm 0.01}$& 2.18$^{\pm 0.04}$& 2.80$^{\pm 0.45}$ & 1.89 &0.71 &26.1 \\
 9&X0501& 09/02& 60189.15& 519& 1.308$^{+0.005}_{-0.005}$&5.92$^{+0.43}_{-0.39}$& 19.1$^{+1.1}_{-1.1}$ & 0.74$^{\pm 0.12}$& 0.39$^{\pm 0.03}$& 2.19$^{\pm 0.03}$& 3.95$^{\pm 0.58}$ & 1.76 &0.47 &26.1 \\
10&X0508& 09/03& 60190.09& 476& 1.446$^{+0.007}_{-0.007}$&5.89$^{+0.51}_{-0.45}$& 18.9$^{+1.3}_{-1.2}$ & 0.77$^{\pm 0.12}$& 0.37$^{\pm 0.02}$& 2.27$^{\pm 0.02}$& 6.78$^{\pm 0.68}$ & 1.45 &0.38 &25.3 \\
11&X0601& 09/04& 60191.13& 512& 1.254$^{+0.006}_{-0.005}$&5.81$^{+0.43}_{-0.38}$& 18.6$^{+1.1}_{-1.0}$ & 0.62$^{\pm 0.12}$& 0.37$^{\pm 0.02}$& 2.20$^{\pm 0.02}$& 6.77$^{\pm 0.58}$ & 1.58 &0.28 &25.1 \\
12&X0608& 09/05& 60192.07& 476& 1.121$^{+0.005}_{-0.005}$&6.32$^{+0.53}_{-0.47}$& 17.6$^{+1.1}_{-1.0}$ & 0.93$^{\pm 0.14}$& 0.36$^{\pm 0.02}$& 2.13$^{\pm 0.03}$& 3.60$^{\pm 0.57}$ & 1.30 &0.39 &24.8 \\
13&X0616& 09/06& 60193.13& 476& 1.478$^{+0.006}_{-0.005}$&6.86$^{+0.47}_{-0.43}$& 17.6$^{+0.9}_{-0.9}$ & 0.84$^{\pm 0.10}$& 0.37$^{\pm 0.01}$& 2.24$^{\pm 0.02}$& 6.13$^{\pm 0.33}$ & 1.49 &0.43 &23.3 \\
14&X0801& 09/07& 60194.10& 502& 1.295$^{+0.008}_{-0.006}$&4.12$^{+0.29}_{-0.27}$& 17.9$^{+0.9}_{-0.9}$ & 0.79$^{\pm 0.12}$& 0.38$^{\pm 0.03}$& 2.18$^{\pm 0.13}$& 6.16$^{\pm 0.40}$ & 1.54 &0.40 &23.2 \\
15&X0901& 09/08& 60195.16& 510& 1.317$^{+0.007}_{-0.006}$&5.77$^{+0.50}_{-0.44}$& 17.3$^{+1.1}_{-1.0}$ & 0.82$^{\pm 0.14}$& 0.36$^{\pm 0.02}$& 2.18$^{\pm 0.03}$& 6.36$^{\pm 0.70}$ & 1.26 &0.35 &22.6 \\
16&X1001& 09/09& 60196.15& 508& 1.384$^{+0.005}_{-0.005}$&7.58$^{+0.68}_{-0.61}$& 16.5$^{+1.1}_{-1.0}$ & 0.69$^{\pm 0.14}$& 0.37$^{\pm 0.02}$& 2.18$^{\pm 0.02}$& 6.90$^{\pm 0.73}$ & 1.26 &0.29 &22.1 \\
17&X1101& 09/10& 60197.15& 505& 1.457$^{+0.005}_{-0.005}$&7.75$^{+0.78}_{-0.68}$& 15.9$^{+1.2}_{-1.1}$ & 0.58$^{\pm 0.14}$& 0.41$^{\pm 0.04}$& 2.14$^{\pm 0.03}$& 4.82$^{\pm 0.67}$ & 1.18 &0.35 &21.3 \\
18&X1201& 09/11& 60198.14& 503& 1.795$^{+0.009}_{-0.008}$&6.18$^{+0.48}_{-0.42}$& 16.3$^{+0.9}_{-0.9}$ & 0.77$^{\pm 0.14}$& 0.37$^{\pm 0.02}$& 2.21$^{\pm 0.03}$& 5.62$^{\pm 0.90}$ & 1.20 &0.35 &20.4 \\
19&X1301& 09/12& 60199.13& 501& 2.349$^{+0.009}_{-0.008}$&7.56$^{+0.60}_{-0.53}$& 16.2$^{+1.0}_{-0.9}$ & 0.83$^{\pm 0.15}$& 0.40$^{\pm 0.04}$& 2.34$^{\pm 0.08}$& 1.47$^{\pm 0.23}$ & 1.13 &0.39 &19.6 \\
20&X1401& 09/13& 60200.12& 500& 2.944$^{+0.027}_{-0.027}$&3.52$^{+0.35}_{-0.31}$& 17.8$^{+1.1}_{-0.8}$ & 0.86$^{\pm 0.12}$& 0.40$^{\pm 0.03}$& 2.35$^{\pm 0.10}$& 0.22$^{\pm 0.04}$ & 1.09 &0.41 &19.6 \\
21&X1402& 09/13& 60200.19& 510& 2.835$^{+0.012}_{-0.011}$&7.92$^{+0.71}_{-0.63}$& 16.2$^{+1.1}_{-1.0}$ & 1.11$^{\pm 0.13}$& 0.37$^{\pm 0.02}$& 2.36$^{\pm 0.07}$& 0.31$^{\pm 0.05}$ & 1.06 &0.54 &19.1 \\
22&X1403& 09/13& 60200.33& 476& 2.619$^{+0.011}_{-0.011}$&5.83$^{+0.38}_{-0.34}$& 16.5$^{+0.8}_{-0.8}$ & 0.92$^{\pm 0.13}$& 0.39$^{\pm 0.03}$& 2.37$^{\pm 0.08}$& 1.04$^{\pm 0.10}$ & 1.07 &0.42 &18.9 \\
23&X1404& 09/13& 60200.46& 476& 2.642$^{+0.009}_{-0.010}$&6.60$^{+0.43}_{-0.40}$& 16.4$^{+0.8}_{-0.8}$ & 1.01$^{\pm 0.15}$& 0.35$^{\pm 0.02}$& 2.43$^{\pm 0.06}$& 1.44$^{\pm 0.16}$ & 1.20 &0.36 &18.8 \\
24&X1405& 09/13& 60200.59& 476& 2.547$^{+0.017}_{-0.017}$&6.32$^{+0.67}_{-0.60}$& 16.1$^{+1.2}_{-1.1}$ & 1.06$^{\pm 0.27}$& 0.32$^{\pm 0.03}$& 2.43$^{\pm 0.10}$& 1.60$^{\pm 0.11}$ & 1.03 &0.33 &18.5 \\
25&X1406& 09/13& 60200.73& 476& 2.157$^{+0.013}_{-0.012}$&6.26$^{+0.69}_{-0.59}$& 16.3$^{+1.3}_{-1.2}$ & 0.80$^{\pm 0.22}$& 0.35$^{\pm 0.03}$& 2.34$^{\pm 0.05}$& 4.87$^{\pm 0.34}$ & 0.93 &0.27 &18.5 \\
26&X1407& 09/13& 60200.86& 476& 2.069$^{+0.008}_{-0.008}$&6.89$^{+0.55}_{-0.48}$& 16.1$^{+0.9}_{-0.9}$ & 1.02$^{\pm 0.14}$& 0.36$^{\pm 0.02}$& 2.34$^{\pm 0.04}$& 4.68$^{\pm 0.92}$ & 1.16 &0.40 &18.2 \\
27&X1408& 09/13& 60200.99& 654& 2.052$^{+0.008}_{-0.008}$&5.82$^{+0.40}_{-0.36}$& 16.3$^{+0.9}_{-0.8}$ & 0.88$^{\pm 0.12}$& 0.37$^{\pm 0.02}$& 2.29$^{\pm 0.03}$& 4.25$^{\pm 0.81}$ & 1.20 &0.38 &18.1 \\
28&X1501& 09/14& 60201.25& 535& 1.924$^{+0.008}_{-0.007}$&6.67$^{+0.56}_{-0.50}$& 16.1$^{+1.0}_{-0.9}$ & 1.97$^{\pm 0.24}$& 0.31$^{\pm 0.01}$& 2.31$^{\pm 0.03}$& 4.51$^{\pm 0.88}$ & 1.05 &0.74 &17.8 \\
29&X1701& 09/16& 60203.23& 536& 2.647$^{+0.015}_{-0.015}$&5.49$^{+0.49}_{-0.43}$& 16.2$^{+1.1}_{-1.0}$ & 0.83$^{\pm 0.13}$& 0.41$^{\pm 0.03}$& 2.32$^{\pm 0.10}$& 0.23$^{\pm 0.05}$ & 1.02 &0.41 &17.6 \\
30&X1801& 09/17& 60204.22& 537& 2.598$^{+0.009}_{-0.008}$&7.94$^{+0.63}_{-0.57}$& 15.4$^{+0.9}_{-0.8}$ & 0.69$^{\pm 0.09}$& 0.44$^{\pm 0.03}$& 2.28$^{\pm 0.06}$& 0.29$^{\pm 0.04}$ & 1.01 &0.39 &16.9 \\
31&X1901& 09/18& 60205.15& 538& 4.353$^{+0.014}_{-0.012}$&7.77$^{+0.58}_{-0.51}$& 15.6$^{+0.9}_{-0.8}$ & 0.64$^{\pm 0.06}$& 1.02$^{\pm 0.04}$& 2.48$^{\pm 0.03}$& 0.16$^{\pm 0.02}$ & 1.02 &0.97 &18.1 \\
32&X2002& 09/19& 60206.21& 524& 7.496$^{+0.052}_{-0.050}$&2.83$^{+0.14}_{-0.13}$& 16.8$^{+0.5}_{-0.5}$ & 0.79$^{\pm 0.04}$& 1.08$^{\pm 0.01}$& 2.61$^{\pm 0.03}$& 0.09$^{\pm 0.02}$ & 1.07 &4.49 &17.2 \\
33&X2003& 09/19& 60206.28& 476& 8.782$^{+0.026}_{-0.026}$&7.30$^{+0.43}_{-0.40}$& 15.0$^{+0.7}_{-0.6}$ & 0.72$^{\pm 0.04}$& 1.09$^{\pm 0.01}$& 2.62$^{\pm 0.03}$& 0.08$^{\pm 0.01}$ & 1.07 &6.56 &15.6 \\
34&X2004& 09/19& 60206.41& 476& 5.592$^{+0.045}_{-0.035}$&7.26$^{+1.81}_{-1.36}$& 10.3$^{+1.5}_{-1.2}$ & 0.84$^{\pm 0.04}$& 1.12$^{\pm 0.02}$& 2.64$^{\pm 0.04}$& 0.17$^{\pm 0.03}$ & 1.03 &3.41 &17.8 \\
35&X2005& 09/19& 60206.54& 476& 5.496$^{+0.023}_{-0.023}$&6.87$^{+0.49}_{-0.46}$& 15.5$^{+0.8}_{-0.7}$ & 0.85$^{\pm 0.06}$& 1.11$^{\pm 0.06}$& 2.64$^{\pm 0.08}$& 0.36$^{\pm 0.09}$ & 0.99 &2.18 &17.9 \\
36&X2101& 09/20& 60207.13& 538& 5.321$^{+0.017}_{-0.018}$&7.28$^{+0.51}_{-0.46}$& 15.3$^{+0.8}_{-0.7}$ & 0.90$^{\pm 0.06}$& 1.06$^{\pm 0.03}$& 2.66$^{\pm 0.03}$& 0.26$^{\pm 0.08}$ & 1.01 &1.70 &17.7 \\
37&X2201& 09/21& 60208.12& 537& 5.246$^{+0.018}_{-0.018}$&7.25$^{+0.54}_{-0.49}$& 15.7$^{+0.8}_{-0.8}$ & 0.84$^{\pm 0.05}$& 1.09$^{\pm 0.03}$& 2.65$^{\pm 0.03}$& 0.27$^{\pm 0.09}$ & 0.97 &1.86 &16.8 \\
38&X2301& 09/22& 60209.18& 537& 3.593$^{+0.016}_{-0.017}$&5.32$^{+0.32}_{-0.29}$& 16.0$^{+0.7}_{-0.7}$ & 0.63$^{\pm 0.45}$& 0.68$^{\pm 0.06}$& 2.46$^{\pm 0.03}$& 0.24$^{\pm 0.07}$ & 1.03 &0.34 &15.3 \\
39&X2401& 09/23& 60210.17& 535& 3.599$^{+0.014}_{-0.013}$&6.33$^{+0.42}_{-0.38}$& 15.6$^{+0.7}_{-0.7}$ & 0.68$^{\pm 0.05}$& 0.58$^{\pm 0.06}$& 2.47$^{\pm 0.03}$& 0.25$^{\pm 0.06}$ & 1.06 &0.31 &14.8 \\
40&X2501& 09/24& 60211.16& 534& 5.836$^{+0.024}_{-0.024}$&5.42$^{+0.36}_{-0.33}$& 16.3$^{+0.7}_{-0.7}$ & 0.82$^{\pm 0.05}$& 1.04$^{\pm 0.02}$& 2.67$^{\pm 0.03}$& 0.27$^{\pm 0.07}$ & 0.99 &2.05 &14.8 \\
41&X2601& 09/25& 60212.15& 532& 6.629$^{+0.033}_{-0.030}$&4.67$^{+0.35}_{-0.33}$& 16.1$^{+0.8}_{-0.7}$ & 0.85$^{\pm 0.05}$& 1.04$^{\pm 0.01}$& 2.73$^{\pm 0.03}$& 0.27$^{\pm 0.05}$ & 0.98 &2.68 &14.2 \\
42&X2701& 09/26& 60213.14& 530& 7.236$^{+0.071}_{-0.071}$&2.42$^{+0.16}_{-0.15}$& 17.9$^{+0.8}_{-0.7}$ & 0.77$^{\pm 0.02}$& 1.02$^{\pm 0.01}$& 2.72$^{\pm 0.03}$& 0.25$^{\pm 0.07}$ & 1.00 &3.36 &13.3 \\
43&X2802& 09/27& 60214.21& 476& 4.985$^{+0.031}_{-0.030}$&4.62$^{+0.34}_{-0.31}$& 16.0$^{+0.8}_{-0.8}$ & 0.90$^{\pm 0.05}$& 0.96$^{\pm 0.03}$& 2.67$^{\pm 0.04}$& 0.43$^{\pm 0.08}$ & 0.99 &0.85 &13.6 \\
44&X2901& 09/28& 60215.12& 524& 3.941$^{+0.017}_{-0.016}$&6.47$^{+0.59}_{-0.53}$& 15.1$^{+0.9}_{-0.8}$ & 0.59$^{\pm 0.02}$& 0.76$^{\pm 0.02}$& 2.50$^{\pm 0.06}$& 0.40$^{\pm 0.09}$ & 0.93 &0.57 &12.6 \\
45&X3001& 09/29& 60216.12& 521& 6.192$^{+0.044}_{-0.043}$&3.78$^{+0.29}_{-0.26}$& 16.9$^{+0.8}_{-0.8}$ & 0.76$^{\pm 0.02}$& 0.97$^{\pm 0.01}$& 2.68$^{\pm 0.04}$& 0.29$^{\pm 0.09}$ & 1.02 &2.24 &12.2 \\
46&X3101& 09/30& 60217.17& 518& 5.403$^{+0.028}_{-0.027}$&5.21$^{+0.39}_{-0.35}$& 15.8$^{+0.8}_{-0.8}$ & 0.78$^{\pm 0.01}$& 0.95$^{\pm 0.01}$& 2.65$^{\pm 0.04}$& 0.33$^{\pm 0.08}$ & 0.94 &1.47 &12.1 \\
47&X3201& 10/01& 60218.16& 515& 6.521$^{+0.041}_{-0.040}$&4.47$^{+0.38}_{-0.33}$& 16.4$^{+0.8}_{-0.8}$ & 0.75$^{\pm 0.02}$& 0.95$^{\pm 0.01}$& 2.68$^{\pm 0.04}$& 0.27$^{\pm 0.05}$ & 0.93 &2.36 &11.3 \\
48&X3301& 10/02& 60219.16& 513& 5.755$^{+0.034}_{-0.035}$&4.30$^{+0.31}_{-0.28}$& 16.7$^{+0.8}_{-0.7}$ & 0.74$^{\pm 0.01}$& 0.94$^{\pm 0.01}$& 2.66$^{\pm 0.04}$& 0.33$^{\pm 0.08}$ & 1.02 &1.69 &11.2 \\
49&X3308& 10/03& 60220.09& 476& 7.077$^{+0.060}_{-0.059}$&3.48$^{+0.33}_{-0.29}$& 16.9$^{+1.0}_{-0.8}$ & 0.66$^{\pm 0.03}$& 0.95$^{\pm 0.01}$& 2.70$^{\pm 0.05}$& 0.27$^{\pm 0.07}$ & 0.96 &2.77 &10.3 \\
50&X3401& 10/04& 60221.14& 509& 8.988$^{+0.057}_{-0.057}$&4.84$^{+0.59}_{-0.49}$& 15.9$^{+1.1}_{-1.1}$ & 0.58$^{\pm 0.04}$& 0.95$^{\pm 0.01}$& 2.70$^{\pm 0.08}$& 0.28$^{\pm 0.03}$ & 1.04 &4.30 &8.52 \\
51&X3502& 10/05& 60222.20& 476& 8.414$^{+0.062}_{-0.063}$&3.95$^{+0.37}_{-0.33}$& 16.4$^{+0.9}_{-0.9}$ & 0.56$^{\pm 0.03}$& 0.96$^{\pm 0.01}$& 2.72$^{\pm 0.07}$& 0.26$^{\pm 0.04}$ & 0.98 &3.62 &8.83 \\

\hline
\end{tabular}
%\noindent{
\leftline{X (=P06143380) is the initial part of ObsID; Q=$FWHM/\nu_{QPO}$; UT in mm/dd format. The units of Exposure (Exp) in s; }
\leftline{$\nu_{QPO}$ in Hz; $N_{\rm H}$ in $10^{22}~{\rm cm}^{-2}$; $T_{in}$ in keV; model fluxes diskbb ($F_{\rm DBB}$) and pexrav ($F_{\rm pexr}$) in $10^{-8}~{\rm erg}~{\rm cm}^{-2}~{\rm s}^{-1}$.}
\end{table*}

\end{document}